\documentclass[conference]{IEEEtran}

\newif\ifFullVersion

\let\labelindent\relax

\usepackage[font={small,it}]{caption}

\usepackage{mathptmx}  
\usepackage{xspace}
\usepackage{graphicx}
\usepackage{psfrag}
\usepackage[hyphens]{url}
\usepackage{pifont}
\usepackage[normalem]{ulem}
\usepackage{color}    
\usepackage{threeparttable}
\usepackage{rotating}
\usepackage{colortbl} 
\usepackage{fancyvrb}
\usepackage{subfigure}
\usepackage{paralist}
\usepackage{comment}
\usepackage{enumitem}
\usepackage{flushend}
\usepackage{multicol}
\usepackage{footmisc}
\usepackage{dblfloatfix}

\usepackage[noadjust]{cite}

\usepackage[usenames,dvipsnames]{xcolor}
\usepackage[breaklinks]{hyperref}
\usepackage{breakurl}

\usepackage{breakurl}

\usepackage{amssymb}
\usepackage{booktabs}
\usepackage{amsmath}

\usepackage{widetext}

\def\centerhack#1{\hbox to 0pt{\hss\footnotesize #1\hss}}

\renewcommand{\paragraph}[1]{\smallskip\noindent\textbf{#1.}}

\newcommand{\name}{SIBRA\xspace}
\newcommand{\nameacr}{Scalable Internet Bandwidth Reservation Architecture\xspace}

\newcommand{\marker}{\textsc{RT}\xspace}

\newcommand\tragedy{tragedy of the network-link commons\xspace}

\newcommand{\src}{source\xspace}
\newcommand{\dst}{destination\xspace}

\newcommand{\rep}{destination\xspace}

\newcommand{\ISD}{ISD\xspace}
\newcommand{\ISDs}{ISDs\xspace}
\newcommand{\AD}{AS\xspace}
\newcommand{\ADs}{ASes\xspace}

\newcommand\bwstd{5}    
\newcommand\bweph{80}   
\newcommand\bwbst{15}   
\newcommand\bwrat{16}   

\newcommand{\footnoteblind}[2]{%
  \addtocounter{footnote}{1}%
  \footnotetext[\thefootnote]{%
    \addtocounter{footnote}{-1}%
    \refstepcounter{footnote}\label{#1}%
      #2%
  }
}

\newlist{ReqR}{enumerate}{1}
\setlist[ReqR]{label=\textbf{Req \arabic*.},leftmargin=*,itemindent=1cm}

\def\centerhack#1{\hbox to 0pt{\hss\footnotesize #1\hss}}
\def\dchack#1{\vbox to 0pt{\vss{\hbox to 0pt{\hss#1\hss}}\vss}}

\title{{\Large \bf \name: \nameacr}}
\begin{document}

\author{
\IEEEauthorblockN{
  Cristina Basescu\IEEEauthorrefmark{1},
  Raphael M. Reischuk\IEEEauthorrefmark{1},
  Pawel Szalachowski\IEEEauthorrefmark{1},
  Adrian Perrig\IEEEauthorrefmark{1},\\
  Yao Zhang\IEEEauthorrefmark{2},
  Hsu-Chun Hsiao\IEEEauthorrefmark{3},
  Ayumu Kubota\IEEEauthorrefmark{4},
  Jumpei Urakawa\IEEEauthorrefmark{4}
}
\IEEEauthorblockA{\IEEEauthorrefmark{1}ETH Zurich, Switzerland}
\IEEEauthorblockA{\IEEEauthorrefmark{2}Beihang University, China}
\IEEEauthorblockA{\IEEEauthorrefmark{3}National Taiwan University, Taiwan}
\IEEEauthorblockA{\IEEEauthorrefmark{4}KDDI R\&D Laboratories Inc., Japan}
}


\IEEEoverridecommandlockouts
\makeatletter\def\@IEEEpubidpullup{9\baselineskip}\makeatother
\IEEEpubid{\parbox{\columnwidth}{Permission to freely reproduce all or part
    of this paper for noncommercial purposes is granted provided that
    copies bear this notice and the full citation on the first
    page. Reproduction for commercial purposes is strictly prohibited
    without the prior written consent of the Internet Society, the
    first-named author (for reproduction of an entire paper only), and
    the author's employer if the paper was prepared within the scope
    of employment.  \\
    NDSS '16, 21-24 February 2016, San Diego, CA, USA\\
    Copyright 2016 Internet Society, ISBN 1-891562-41-X\\
    http://dx.doi.org/10.14722/ndss.2016.23132
}
\hspace{\columnsep}\makebox[\columnwidth]{}}

\maketitle

\begin{abstract}
  This paper proposes a \nameacr (\name) as a new approach against DDoS attacks,
which, until now, continue to be a menace on today's Internet. \name provides
scalable inter-domain resource allocations and \textit{botnet-size
independence}, an important property to realize why previous defense approaches
are insufficient. Botnet-size independence enables two end hosts to set up
communication regardless of the size of distributed botnets in any Autonomous
System in the Internet. \name thus ends the arms race between DDoS attackers
and defenders. Furthermore, \name is based on purely stateless operations for
reservation renewal, flow monitoring, and policing, resulting in highly
efficient router operation, which is demonstrated with a full implementation.
Finally, \name supports Dynamic Interdomain Leased Lines (DILLs), offering new
business opportunities for ISPs.

\end{abstract}

\section{Introduction}
\label{sec:introduction}\label{sec:intro}
\noindent A recent extensive discussion among network administrators on the
NANOG mailing list~\cite{website:nanog} pointedly reflects the current state of
DDoS attacks and the trickiness of suitable defenses: defenses typically
perform traffic scrubbing in ISPs or in the cloud, but attacks often surpassing
20--40 Gbps overwhelm the upstream link bandwidth and cause congestion that
traffic scrubbing cannot handle. As attacks of up to 400 Gbps have recently
been observed~\cite{website:400}, no vital solution seems to be on the horizon
that can defend the network against such large-scale flooding attacks.

\emph{Quality of service} (QoS) architectures at different granularities, such
as IntServ~\cite{intserv} and DiffServ~\cite{diffserv}, fail to provide
end-to-end traffic guarantees at Internet scale: with billions of flows through
the network core, routers cannot handle the per-flow state required by IntServ,
whereas the behavior of DiffServ's traffic classification across different
domains cannot guarantee consistent end-to-end connectivity.

\emph{Network capabilities} \cite{Anderson2004,Yaar2004,yang2005tva,natu2007fine,lee2010floc} 
are not effective against attacks such as
Coremelt~\cite{studer2009coremelt} that build on \textit{legitimate}
low-bandwidth flows to swamp core network links. FLoc~\cite{lee2010floc} in
particular considers bot-contaminated domains, but it is ineffective in case of
dispersed botnets.

\emph{Fair resource reservation} mechanisms
(\textit{per source}~\cite{rfc970},
\textit{per destination}~\cite{yang2005tva},
\textit{per flow}~\cite{demers1989fair,Yaar2004,intserv},
\textit{per computation}~\cite{Parno2007},
and \textit{per class}~\cite{diffserv})
are necessary to resolve link-flooding attacks, but are not sufficient:
none of them provides \textit{botnet-size independence}, a critical property
for viable DDoS defense.

\textbf{Botnet-size independence} is the property in which a legitimate flow's
allocated bandwidth does not diminish below the minimum allocation when the
number of bots (even in other ASes in the world) increases. Per-flow and
per-computation resource allocation, for instance, will reduce their allocated
bandwidth towards 0 when the number of bots that share the corresponding
resources increases.

To illustrate the importance of botnet-size independence, we observe how
previous systems suffer from the \textit{\tragedy}, which refers to the problem
that the allocation of a shared resource will diminish toward an
infinitesimally small allocation when many entities have the incentive to
increase their ``fair share''.\footnote{We use this term following Garrett
Hardin's \emph{Tragedy of the Commons} \cite{hardin1968tragedy}, which
according to the author has no technical solution, but instead ``requires a
fundamental extension in morality''. As we should not expect attackers to show
any of the latter, we believe in a technical solution --- at least for the
Internet!}
In particular, per-flow fair sharing allocations (including per-class
categorization of flows) suffer from this fate, as each source has an incentive
to increase its share by simply creating more flows. However, even when the
fair sharing system is not abused, the resulting allocations are too small to
be useful. To explain in more detail, denoting $N$ as the number of end hosts
in the Internet, per-source or per-destination schemes could ideally conduct
fair sharing of $O(1/N)$ based on all potential sources or destinations that
traverse a given link. However, with increasing hop-count distance of the link
from the source or to the destination, the number of potential sources or
destinations that traverse that link increases exponentially.
Per-flow reservation performs even more poorly, allocating a bandwidth slice of
only $O(1/M^2)$ in the case of a Coremelt attack~\cite{studer2009coremelt}
between $M$ bots, and only $O(1/M\!*\!P)$ during a Crossfire
attack~\cite{kang2013crossfire} with $P$ destination servers that can be
contacted. In the presence of \textit{billions} of end hosts engaged in
end-to-end communication, the allocated bandwidth becomes too small to be
useful.

In this paper, we propose a \nameacr (\name), a novel bandwidth
allocation system that operates at Internet-scale and resolves the
drawbacks of prior systems. In a nutshell, \name provides interdomain
bandwidth allocations, which enable construction of Dynamic
Interdomain Leased Lines (DILLs), in turn enabling new ISP business
models. \name's bandwidth reservations guarantee a minimal amount of
bandwidth to each pair of end hosts by limiting the possible paths in end-to-end
communication. An important property of \name is its per-flow
stateless operation for reservation renewal, monitoring, and 
policing, which results in scalable and efficient router operation.
\name is fully implemented; our evaluation demonstrates its effectiveness.

\section{Goals, Assumptions, and the Adversary}
\label{sec:probdef}
\noindent The goal of this paper is to defend against \textit{link-flooding
attacks}, in which distributed attackers collude by sending traffic to each
other (Coremelt~\cite{studer2009coremelt}) or to publicly accessible servers
(Crossfire~\cite{kang2013crossfire}) in order to exhaust the bandwidth of
targeted servers and Internet backbone links. In the case of Coremelt, the
traffic volume might not be limited (e.g., by TCP congestion control) since all
participating hosts are under \emph{adversarial} control and can thus run any
protocol. In the case of Crossfire, distributed attackers collude by sending
traffic to \emph{legitimate} hosts in order to cut off network connections to
selected servers.  We note that other known attacks constitute a combination of
the two cases above.

\paragraph{Adversary model} We assume that \ADs may be malicious and misbehave
by sending large amounts of traffic (bandwidth requests and data packets). We
furthermore assume any \AD in the world can contain malicious end hosts (e.g.,
as parts of larger botnets). In particular, there is no constraint on the
distribution of compromised end hosts.
However, attacks launched by routers (located inside ASes) that intentionally
modify, delay, or drop traffic (beyond the natural drop rate) are out of the
scope of this paper.

\paragraph{Desired properties}
Under the defined adversary model, we postulate the following properties a
\textit{link-flooding-resilient bandwidth reservation mechanism} should
satisfy:

\begin{compactitem}

\item \textbf{Botnet-size independence.} The minimum amount of guaranteed
bandwidth per end host does not diminish with an increasing number of bots. 

\item \textbf{Per-flow stateless operation.} The mechanism's overhead on
routers should be negligible. In particular, backbone routers should not
require per-flow, per-source, or per-destination state in the fastpath, which
could lead to state exhaustion attacks.\footnote{A router's \textit{fastpath}
handles packet processing and forwarding on the line card, and is thus
performance-critical. Routing protocols, network management, and flow setup are
handled by the \textit{slowpath}, which typically executes on the main router
CPU and is thus less performance-critical.} Our analysis of real packet traces
on core links supports this property (Section~\ref{sec:stateless}).

\item \textbf{Scalability.} The costs and overhead of the system should scale to
  the size of the Internet, including management and setup, \AD
  contracts, router and end host computation and memory, as well as
  communication bandwidth.

\end{compactitem}

\paragraph{Network assumptions} To achieve the properties we seek, we
assume (i) a network architecture that provides source-controllable network
paths, and (ii) hierarchical bandwidth decomposition.

Concerning the first assumption of \emph{source-controllable network paths},
we assume that routing paths (i.e., sequences of \AD hops) are selected from
several options by bandwidth-requesting sources (who then negotiate bandwidth
with the destination and the intermediate \AD{} hops). There are multiple
routing protocols that provide such features: Pathlet
routing~\cite{godfrey2009pathlet}, NIRA~\cite{yang2007nira}, and
SCION~\cite{zhang2011scion,scion2015}, where the source can specify a path in
the packet headers, or I3~\cite{stoica2002internet} and
Platypus~\cite{raghavan2009secure}, where the source specifies a sequence of
forwarding nodes. We note that this first assumption may be of independent
interest for ISPs since they may financially benefit
\cite{LasJohChu08:userDirectedRouting}.

Our second assumption of \emph{bandwidth decomposition} is satisfied through a
concept of \emph{domain isolation}. To this end, we leverage SCION's isolation
concept \cite{zhang2011scion,scion2015} by grouping ASes into independent
\emph{Isolation Domains} (ISDs), each with an isolated control plane.
\autoref{fig:topology} depicts an example of 4 ISDs. The two end hosts
$S$ and $D$ in \emph{different} ISDs are connected by stitching three types of
path segments together: an \emph{up-segment} from $S$ to its ISD core, a
\emph{core-segment} within the Internet core (from source ISD to destination
ISD), and a \emph{down-segment} from $D$'s ISD core to end host $D$. The ISD
\textit{core} refers to a set of top-tier \ADs, the \emph{core \ADs}, that
manage the ISD (depicted with a dark background in
\autoref{fig:topology}).
Intuitively, the isolation property yields that \ADs inside an ISD can
establish paths with bandwidth guarantees to the ISD core --- independently of
bandwidth reservations in other ISDs. The bandwidth reservations for paths
\emph{across} ISDs will then be based on the reservations \emph{inside} the
ISDs, but will be lower- and upper-bounded for each end host. In particular,
malicious entities will not be able to congest the network.

Furthermore, we assume that each end-to-end flow from $S$ to $D$ can be
assigned a
unique, non-hijackable \textit{flow identifier}~\cite{rfc5201,rfc6253,AIP};
that ASes locally allocate resources to their internal end hosts;
and that network links can fail and exhibit natural packet loss, which could lead to
dropped reservation requests or dropped data packets.

\section{\name Design}
\label{sec:path}
\noindent This section describes the design of \name, in particular bandwidth
reservations and their enforcement. After a brief overview, we describe \name's
reservation types in detail.

\subsection{SIBRA overview}
\label{sec:overview}

\noindent A key insight of \name is its hierarchical decomposition of the
bandwidth allocation problem to make management and configuration scale to the
size of the Internet. More specifically, \name makes use of 
(1)~\textbf{core} contracts:
\emph{long-term} contracts amongst the core ASes of large-scale isolation domains (ISDs),
(2)~\textbf{steady} contracts:
\emph{intermediate-term} contracts amongst \ADs within an ISD, and
(3)~\textbf{ephemeral} contracts:
\emph{short-term} contracts for end-to-end communication that leverage the long-term
and inter\-mediate-term contracts.

\begin{figure}[t]
  \begin{center}
    \includegraphics[width=0.9\linewidth]{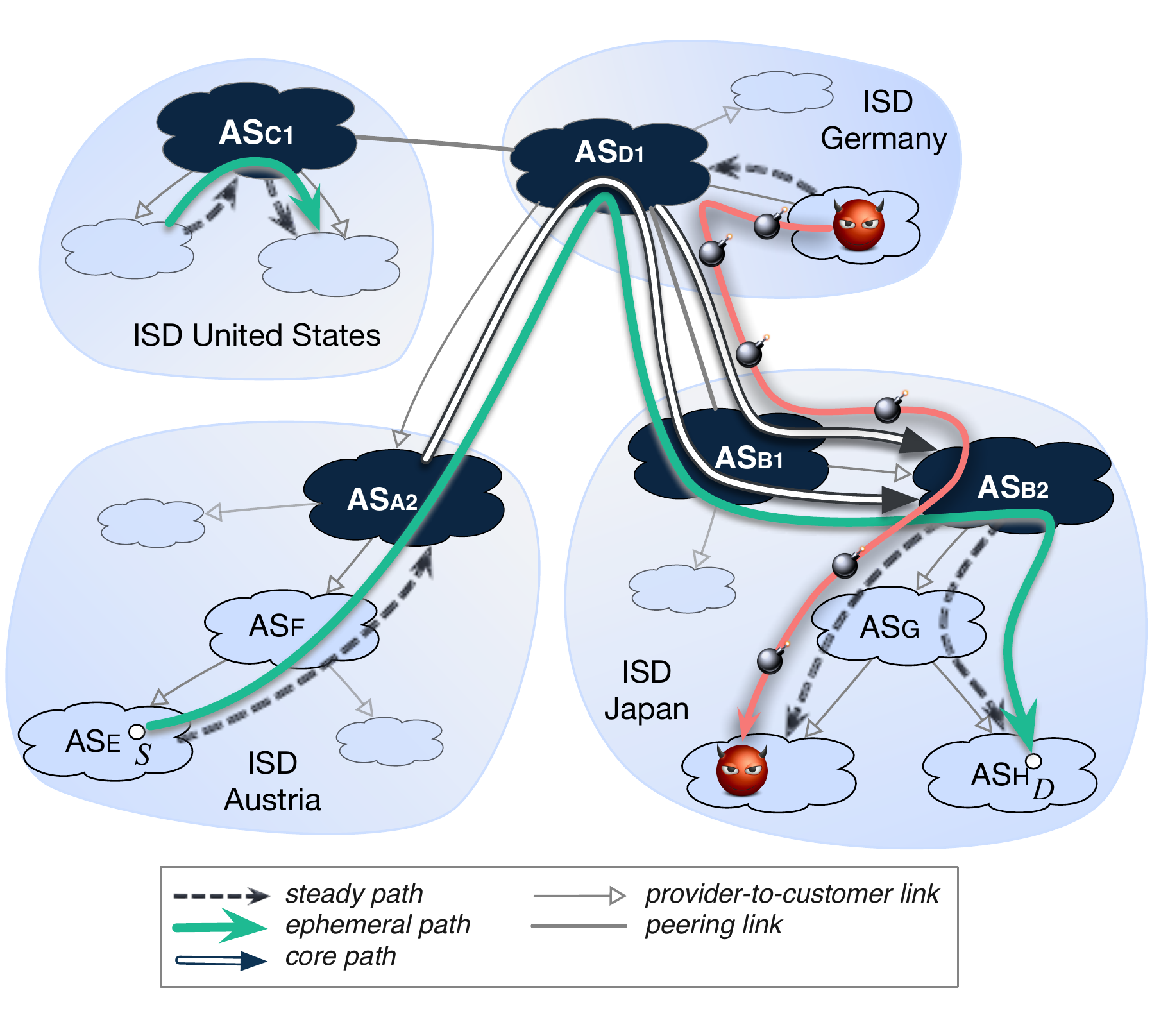}
  \end{center}
  \vspace{-5mm}
  \caption{Exemplary \name topology with 4 isolation domains and their ASes
  (the core ASes are filled). The ephemeral path (green) from end host $S$ to
  end host $D$ is created along a steady up-path, a core path, and a steady
  down-path. The attack traffic (red) does not diminish the reserved bandwidth
  on ephemeral paths.}
  \label{fig:topology}
  \vspace{-7mm}
\end{figure}

Thanks to this three-layer decomposition, on the order of 100 large-scale ISDs (e.g.,
composed by sets of countries or groups of companies) can scalably establish
long-term \textbf{core~paths} with guaranteed bandwidth between each other (the
double continuous lines in \autoref{fig:topology}). Within each ISD, providers
sell bandwidth to their customers, and customers can establish
inter\-mediate-term reservations for specific \emph{intra}-ISD paths, which we
call \textbf{steady paths} (the dashed lines in \autoref{fig:topology}). Steady
paths are mostly used for connection setup traffic, but can also be used for
low-bandwidth data traffic.
Finally, core and steady paths in conjunction enable the creation of short-term
end-to-end reservations \emph{across} ISDs, which we call
\textbf{ephemeral paths} (the solid green lines in
\autoref{fig:topology}). Ephemeral paths, in contrast to steady paths,
are used for the transmission of high-throughput data traffic.

\begin{figure}[t]
  \begin{center}
    \includegraphics[width=0.7\linewidth]{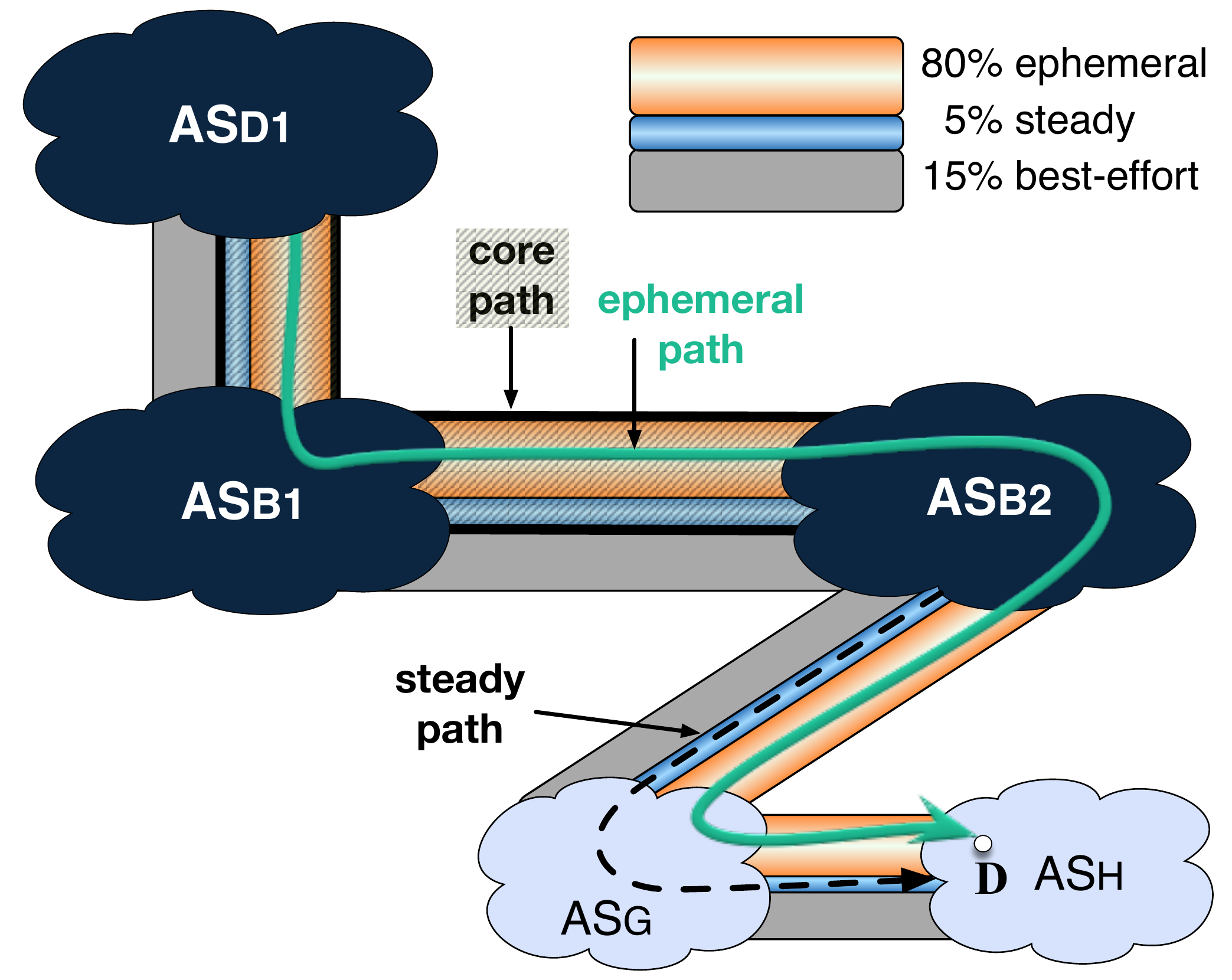}
  \end{center}
  \vspace{-3mm}
  \caption{The anatomy of \name links: \bweph\% of the link bandwidth is used
  for ephemeral traffic, \bwstd\% for steady traffic, and \bwbst\% for
  best-effort traffic. The core path from $AS_{D1}$ to $AS_{B2}$ comprises
  steady and ephemeral traffic, but excludes best-effort traffic.}
  \vspace{-7mm}
  \label{fig:anatomy}
\end{figure}

\name paths are established over \name links whose anatomy
is depicted in \autoref{fig:anatomy}: \bweph\% of
the bandwidth of each \name link is allocated for ephemeral traffic, \bwstd\% for
steady traffic, and the remaining \bwbst\% for best-effort traffic.
These proportions are flexible system parameters; we discuss the current choice
in \autoref{sec:parameters}. Note that the proportion for steady and ephemeral
traffic constitutes an upper bound: in case the ephemeral bandwidth is not
fully utilized, it is allocated to best-effort traffic
(Section~\ref{sec:ephemeral}).

An important feature of \name is that steady paths, besides carrying the
\bwstd\% control traffic of links inside an ISD, also limit the bandwidth for ephemeral
paths: An ephemeral path is created by launching a request through existing
steady paths whose amounts of bandwidth determine -- up to a fixed scaling factor
-- the bandwidth of the requested ephemeral path. More precisely, an ephemeral
path is created through the combination of (i) a steady up-path in the source
ISD, (ii) the steady part of a core path, and (iii) a steady down-path in the
destination ISD.\footnote{For instance, \autoref{fig:topology} shows an
ephemeral path from host $S$ in $AS_E$ to host $D$ in $AS_H$. If the source and
destination are in the same ISD, then the core path may not be necessary, e.g.,
the ephemeral path inside the US ISD.} The ephemeral path \emph{request} uses
only the steady portion of a link (the blue part in \autoref{fig:anatomy}); the
actual ephemeral path \emph{traffic} uses only the ephemeral portion of a link
(the orange part in \autoref{fig:anatomy}). In other words, the more steady
bandwidth a customer purchases locally within her ISD, the larger the fraction
of ephemeral bandwidth she obtains to any other ISD in the Internet.

Based on these ideas, it becomes intuitively clear how botnet-size independence
is achieved and how the \tragedy is resolved: Each pair of domains can obtain a
minimum bandwidth allocation, based on their respective steady paths and based
on the core contract. Thus, a botnet cannot influence the minimum allocation,
no matter its size and distribution. A bot can only use up the bandwidth
allocated to the AS it resides in, but not lower the minimum allocation of any
other AS. It is thus in the responsibility of an AS to manage its allocations,
and thereby to prevent bots from obtaining resources of others within that AS.

In case an \AD is dissatisfied with its minimum allocation, it can purchase
more bandwidth for its steady paths, as well as request its core \AD to
purchase a larger allocation for the core contract, which the \AD would likely
need to pay for. An important point of these contracts is that, in order to
scale, core contracts are purely neighbor-based: only neighboring \ADs perform
negotiations.

\name's scalability is additionally based on a relatively low number of
ephemeral paths, compared to \emph{all possible} end-to-end paths in today's
Internet, considered for instance by IntServ~\cite{intserv}. As mentioned
above, an ephemeral path in \name is fully determined by choosing two steady
paths and a core path. The number of steady up-/down-paths an \AD can
simultaneously have is upper-bounded by a small \name system parameter (e.g.,
$5$ to $7$), and the number of core paths is naturally upper-bounded by the
number of ISDs.

To make \name viable for practical applications, we need to ensure that all
aspects of the system are scalable and efficient, which holds in particular for
the frequent operations such as flow admission, reservation renewal, and
monitoring and policing.
For instance, all fastpath operations are per-flow stateless to avoid
state-exhaustion attacks and to simplify the router architecture.

\subsection{Core paths}
\label{sec:corepaths}

\noindent
Directly-connected core \ADs (i.e., Tier-1 ISPs) are expected to agree on a
guaranteed amount of bandwidth for traffic traversing their connecting links.
We envision that \ADs ratify such \emph{core contracts} on mutual business
advantages for their customers, on top of currently negotiated
customer-to-provider or peering relations. Similar to SLAs, core contracts are
long term (e.g., on the order of months) and can have an up-time associated
(e.g., the bandwidth is guaranteed 99.99\% of the time). Core contracts
comprise steady and ephemeral traffic, as illustrated in the shaded part of
\autoref{fig:anatomy}.
If one of the \ADs sends more traffic than agreed on, the \AD is held
accountable, according to the established contract.

\begin{figure}[t]
  \begin{center}
    \includegraphics[width=0.95\linewidth, trim=0 30 0 0]{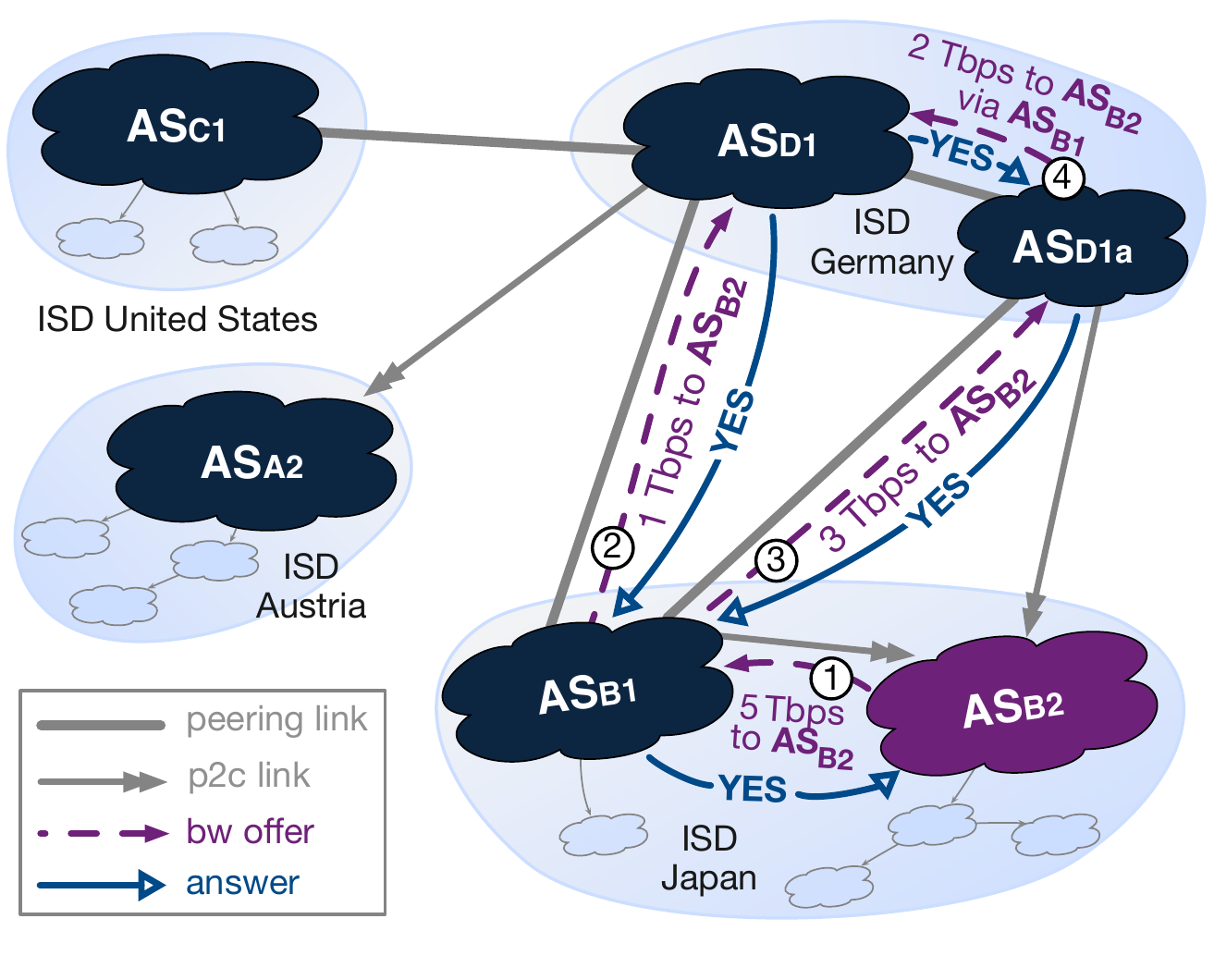}
  \end{center}
  \caption{Core contracts between core \ADs ($AS_{D1}$, $AS_{D1a}$, $AS_{B1}$,
    $AS_{B2}$).}
  \vspace{-3mm}
  \label{fig:corecontracts}
\end{figure}

Core contracts are initiated by \textit{receiver core \ADs}: each core \AD
observes the historical traffic values \textit{received} on its neighboring
links, and proposes in the core contracts similar values for the traffic the
\AD is willing to absorb. For instance, in \autoref{fig:corecontracts},
$AS_{B2}$ proposes to absorb 5~Tbps of steady and ephemeral traffic from
$AS_{B1}$ (Step~\ding{172}), and $AS_{B1}$ accepts. The contract is followed as long as $AS_{B1}$
sends at most 5~Tbps to $AS_{B2}$, regardless of whether $AS_{B1}$ is the
actual origin of the traffic, or $AS_{B1}$ only forwards someone else's traffic
to $AS_{B2}$. For instance, $AS_{B1}$ could forward traffic from $AS_{D1}$ and
$AS_{D1a}$ to $AS_{B2}$. In the example, $AS_{B1}$ offers to forward 1~Tbps
from $AS_{D1}$ (Step~\ding{173}), and 3~Tbps from $AS_{D1a}$
(Step~\ding{174}). $AS_{D1a}$ extends the latter contract by proposing to
$AS_{D1}$ to absorb 2~Tbps towards $AS_{B2}$ (Step~\ding{175}). After
completion of the negotiation, $AS_{D1}$ obtains guaranteed bandwidth to
$AS_{B2}$ along two \textbf{core paths}.

\label{sec:contracts}

\autoref{tab:core} illustrates a local \emph{guaranteed-bandwidth} table that
stores such core paths for $AS_{D1}$. The table resembles a forwarding table
and may contain multiple entries for each destination core \AD, one entry for
each core path. It results from the contract proposals and the received
acknowledgments for a specific destination, $AS_{B2}$ in this case. For
brevity's sake, \autoref{tab:core} shows only the entries for destination
$AS_{B2}$.

\begin{figure}[tp]
\small
\renewcommand{\arraystretch}{1.0}
\centering
 \begin{tabular}{lll}
 \hline
 Destination & \centering Path & Bandwidth \\ [0.5ex]
 \hline\hline
 $AS_{B2}$& $AS_{B1} \rightarrow AS_{B2}$ & 1 Tbps\\
 $AS_{B2}$& $AS_{D1a} \rightarrow AS_{B1}\rightarrow AS_{B2}$ & 2 Tbps \\ [-1ex]
 ~ $\vdots$ & ~ $\vdots$ & ~ $\vdots$ \\
 \hline
 \end{tabular}
\caption{Core contracts table at $AS_{D1}$. Two core paths lead to $AS_{B2}$.}
\vspace{-5mm}
\label{tab:core}
\end{figure}

The bandwidth of a core path reflects the overall traffic volume exchanged
between the source and the destination ASes. To bootstrap the process, each
participating AS observes aggregate traffic volumes on its neighboring links,
and initiates contracts with a bandwidth of 85\% of the observed aggregate
volume (5\% steady + 80\% ephemeral). The initially estimated contracts are
refined as dictated by the customer requirements and payments (explained below).

\paragraph{Scalability} The core contract proposals traverse only one link
before being accepted or denied. For instance, in \autoref{fig:corecontracts},
$AS_{B1}$ first accepts $AS_{B2}$'s proposal (Step~\ding{172}), and only
afterwards, it submits its offers (Steps~\ding{173} and \ding{174}). Achieving
global consensus through immediate agreements is possible due to the
\textit{destination-initiated} process of establishing core contracts, in which
the supported amount of traffic is already specified and can thus be decided
based on \textit{local knowledge}. In contrast, source-initiated requests would
require a distributed consensus algorithm that would traverse all ASes whose
agreement is required. \name's design decision sacrifices such costly
interactions for better scalability, achieving a core contract design that is
scalable with the number of core \ADs.

\paragraph{Payment} Core paths not only guarantee bandwidth between
ISDs, they also regulate the traffic-related money flow between core
\ADs according to existing provider-to-customer (p2c) or peering (p2p)
relationships (e.g., c2p between $AS_{B2}$ and $AS_{B1}$, and p2p
between $AS_{D1}$ and $AS_{B1}$).

Similar to today's state of affairs, we believe that market forces create a
convergence of allocations and prices when ASes balance the bandwidth between
their peers and adjust the contracts such that the direct core \AD neighbors
are satisfied. The neighbors, in turn, recursively adapt their contracts to
satisfy the bandwidth requirements of their customers. Paying customers thus
indirectly dictate to core \ADs the destination ISDs of core paths and the
specified bandwidth in the contracts.

\subsection{Steady paths}

\noindent Steady paths are intermediate-term, low-bandwidth
reservations that are established by ASes for guaranteed
communication availability within an ISD. We envision that the
default validity of steady paths is on the order of minutes, but it
can periodically be extended. An endpoint \AD can voluntarily tear
down its steady path before expiration and set up a new steady path.
For example, in \autoref{fig:topology}, $AS_E$ sets up a steady path
to $AS_{A2}$, and $AS_H$ requests bandwidth guarantees from
$AS_{B2}$. As mentioned earlier, \name uses steady paths for two
purposes: (1)~as communication guarantees for low-bandwidth
traffic, and (2)~as building block for ephemeral paths: to
guarantee availability during connection setup and to perform
weighted bandwidth reservations (Section~\ref{sec:ephemeral}).

\begin{figure}[t]
  \begin{center}
  \includegraphics[width=0.95\linewidth, trim=0 25 0 0]{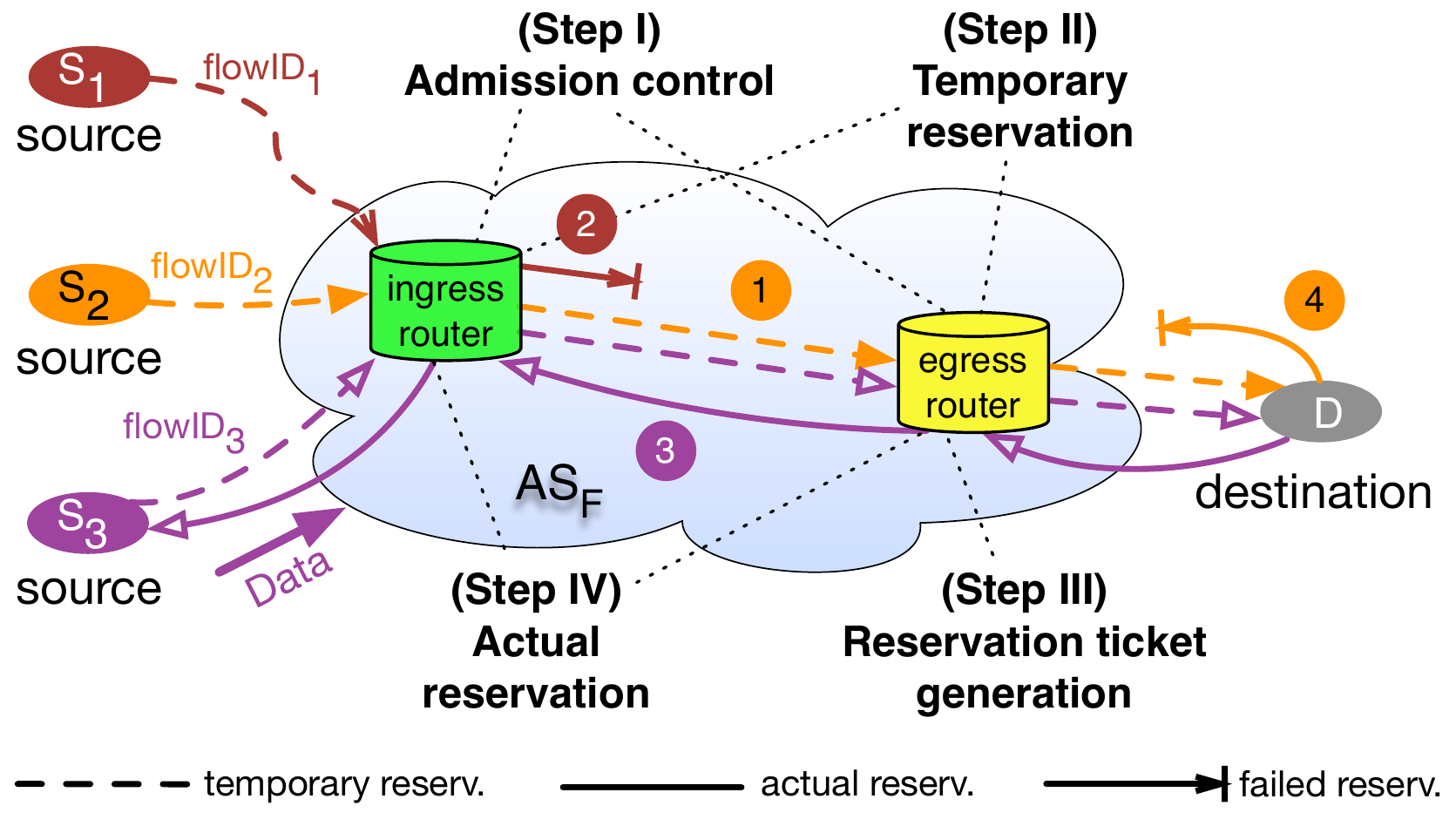}
  \end{center}
  \caption{Transit $AS_F$ processing reservation requests for sources $S_1$, $S_2$, $S_3$ and
  destination $D$.\footref{fnt:destination}}
  \vspace{-5mm}
  \label{fig:reservation}
\end{figure}

\footnoteblind{fnt:destination}{We use the term \emph{destination} in the following (and also in
\autoref{fig:reservation}) to stay as general as possible. For steady-path
reservation requests, the destination is the ISD core; for ephemeral-path
reservation requests, the destination will be another end host
(\autoref{sec:ephemeral}).}

\paragraph{Reservation request}
\name leverages so-called SCION \emph{routing beacons}~\cite{scion2015} that
disseminate top-down from the ISD core to the \ADs. On their journey down, they
collect AS-level path information as well as information about the current
amount of available bandwidth (both steady and ephemeral) for each link.
When a leaf \AD receives such a routing beacon with information about
a path segment, the \AD can decide to submit a reservation request
that promotes the path segment to a steady path. In this case, the
leaf \AD (e.g., $AS_E$ in \autoref{fig:topology}, or $S_3$ in
\autoref{fig:reservation}) computes a new flow ID, chooses the amount
of bandwidth and the expiration time, and sends a \emph{steady path
reservation message} up the path to the core. 
The requested amount of bandwidth can be chosen from a number of predefined
bandwidth classes, introduced for monitoring optimization purposes
(Section~\ref{sec:mon}).

Each intermediate AS on the path to the core performs \textit{admission
control} by verifying the availability of steady bandwidth to its
neighbors on the path (Step~I in \autoref{fig:reservation}). Given
the fact that inbound traffic from multiple ingress routers may
converge at a single egress router, admission control is performed at
both ingress and egress routers. Specifically, the ingress router of
$AS_i$ checks the availability of steady bandwidth on the link
$AS_{i-1}\rightarrow\,AS_i$, and the egress router of $AS_i$ on the
link $AS_i\rightarrow\,AS_{i+1}$.
If enough bandwidth is available at both the ingress and the egress
router (Case~\ding{202} in \autoref{fig:reservation}), both routers
\textit{temporarily} reserve the requested bandwidth (Step~II).
Subsequently, the egress router of $AS_i$ issues a cryptographically
authenticated \textit{reservation token} (RT) encoding the positive
admission decision (Step III).

An RT generated by $AS_i$ is authenticated using a cryptographic key
$K_i$ known only to $AS_i$, by which $AS_i$ can later verify if an RT
embedded in the data packet is authentic.
More specifically, the RT contains the authenticated ingress and the
egress interfaces of $AS_i$, and the reservation request information.
RTs are onion-authenticated to prevent an attacker from crafting a
steady path from RT chunks:

\vspace{-12pt}
\begin{equation*}
\begin{aligned}
\quad & \mathit{RT}_\mathit{AS_i} =
  \mathit{ingress_{AS_i}} \,\parallel\, \mathit{egress_{AS_i}} \,\parallel \\
\quad & \quad
  MAC_{K_i}\big( \mathit{ingress_{AS_i}} \parallel \mathit{egress_{AS_i}} \parallel \mathit{Request} \parallel RT_\mathit{AS_{i-1}}\big)
\end{aligned}
\label{eq:RT}
\end{equation*}
where $\mathit{Request}$ is defined as
$\mathit{Bw_{req}}\parallel\mathit{ExpTime}\parallel\mathit{flowID}$.
We emphasize that steady path reservation flow identifiers are independent of
TCP flow identifiers: A steady path can carry packets from \emph{multiple} TCP
flows, as long as these packets contain the RTs corresponding to the steady
path in their header.

If at least one of the routers of $AS_i$ cannot meet the request
(Case \ding{203}), it suggests an amount of bandwidth that could be
offered instead, and adds this suggestion to the packet header.
Although already failed, the request is still forwarded to the
destination (i.e., to the ISD core in case of steady paths) to collect
suggested amounts of bandwidth from subsequent
\ADs. This information helps the source make an informed and direct
decision in a potential bandwidth re-negotiation.

As steady paths are only infrequently updated, scalability and
efficiency of steady path updates are of secondary importance.
However, $AS_i$ can still perform an efficient admission decision by
simply considering the current utilization of its directly adjacent
\AD neighbors.
Such an efficient mechanism is necessary for reservation requests (and
renewals) to be fastpath operations, avoiding to access per-path state.
In case of a positive admission decision, $AS_i$ needs to account for
the steady path individually per leaf \AD where the reservation
originates from. Only slowpath operations, such as policing of
misbehaving steady paths, need to access this per-path information
about individual steady paths.

\paragraph{Confirmation and usage} When the reservation request reaches the
destination~$D$, the destination replies to the requesting source (e.g., $S_3$)
either by a \textit{confirmation message} (Case \ding{204} in
\autoref{fig:reservation}) containing the RTs accumulated in the request packet
header, or by a \textit{rejection message} (Case \ding{205}) containing the
suggested bandwidth information collected before.\footref{fnt:destination} As
the confirmation message travels back to the source, every ingress and egress
router accepts the reservation request and switches the reservation status from
temporary to active (Step IV).

In order to use the reserved bandwidth for actual data traffic, the
source includes the RTs in the packet header.

\subsection{Ephemeral paths}
\label{sec:ephemeral}

\noindent Ephemeral paths are used for communication with guaranteed high
bandwidth. Ephemeral paths are short-lived, only valid on the order of tens of
seconds, and thus require continuous renewals through the life of the connection.
The source, the destination, and any on-path \AD can rapidly renegotiate the
allocations. \autoref{fig:topology} shows two ephemeral paths, one inside an
ISD, one across three ISDs.

We emphasize that the amount of ephemeral bandwidth that is proportional to
steady bandwidth may constitute a lower bound: If more ephemeral bandwidth is
available (for instance since not everybody might be using his fair share of
ephemeral bandwidth), requesters can choose a bandwidth class \emph{above} the
proportional ratio. In the spirit of fair allocation of joint resources, the
lifetime of ephemeral paths is limited to 16 seconds in order to curtail the
time of resource over-allocation. The details of the over-allocation, however,
are out of scope and left for future work.

\paragraph{Ephemeral paths from steady paths} Ephemeral path requests bear many
similarities with steady path requests, yet bootstrapping is different: An
ephemeral path reservation is launched by an end host, as opposed to a steady
path reservation that is launched by a leaf \AD. The end host (e.g., host $S$ in
\autoref{fig:topology}) first obtains a steady up-path starting at its \AD
(e.g., $AS_E$) to the ISD core, and a steady down-path starting at the
destination ISD core (e.g., $AS_{B2}$) to the destination leaf \AD (e.g.,
$AS_H$).  Joining these steady paths with an inter-ISD core path (e.g., from
$AS_{A2}$ to $AS_{B2}$) results in an end-to-end path $P$, which is used to
send the ephemeral path request from the source end host $S$ to the destination
end host $D$ using allocated steady bandwidth.

More specifically, $S$ first generates a new flow ID, chooses an amount of
bandwidth to request from \name's predefined ephemeral bandwidth classes, and
sends the ephemeral path request along path $P$.\footnote{Similarly to the
steady path case, although an ephemeral path is identified by a flow ID, this
flow ID is orthogonal to TCP flow IDs. A single ephemeral path can transport any
data packets regardless of their layer-4 protocol.} Recall that the path is
composed of a steady up-path of $S$, a core path, and a steady down-path of
$D$.  The leaf \AD where the source end host resides (e.g., $AS_E$) may decide
to block the request in some cases, for instance if the bandwidth purchased by
the leaf \AD is insufficient. Each intermediate \AD on path $P$ performs admission
control through a weighted fair sharing mechanism that ensures the ephemeral
bandwidth is directly proportional with its steady path bandwidth, as described
next. The bandwidth reservation continues similarly to the steady path case.

If bots infest source and destination leaf \ADs, these bots may try to exceed
their fair share by requesting, respectively approving, excessively large
amounts of bandwidth. To thwart this attack, each leaf \AD is responsible for
splitting its purchased bandwidth among its end hosts according to its local
policy, and for subsequently monitoring the usage.

\paragraph{Efficient weighted bandwidth fair sharing} The intuition behind
\name's weighted fair sharing for ephemeral bandwidth is that purchasing steady
bandwidth (or generally spoken: bandwidth for control traffic) on a link $L$
guarantees a proportional amount of ephemeral bandwidth on $L$. In
\autoref{fig:topology}, the ephemeral bandwidth on the ephemeral path from end
host $S$ to $D$ is proportional to the steady bandwidth on the steady up-path
from $AS_E$ to core $AS_{A2}$, and also proportional to the steady bandwidth on
the steady down-path from core $AS_{B2}$ down to $AS_H$.
We explain the details of the three cases of intra-source-ISD links, core
links, and intra-destination-ISD links in the following.

\emph{(1) Ephemeral bandwidth in the source ISD.}
For instance, a steady up-path of $500$ kbps traversing intra-ISD
link $L$ guarantees $\frac{\bweph}{\bwstd}\cdot500$ kbps of ephemeral
bandwidth on $L$. Note that $\frac{\bweph}{\bwstd}=\bwrat$ is the
ratio between ephemeral and steady bandwidth
(\autoref{sec:overview}). Generally speaking, a steady up-path
$S_u$ with steady bandwidth $sBW_u$ traversing $L$ can request
ephemeral bandwidth of
\begin{equation}
  \label{eq:eph:source}
  eBW_u = \bwrat \cdot sBW_u
\end{equation}
Consequently, an \AD that purchases a steady up-path $S_u$ can
guarantee its customers a fixed amount of ephemeral bandwidth for
customers' ephemeral path requests launched via $S_u$, regardless of
the ephemeral path requests from other \ADs on $L$.

To provide bandwidth guarantees on \emph{every} link to a destination,
\name extends the influence of the purchased steady up-path
bandwidth along the path to the destination AS. In fact, \name's
weighted fair sharing for ephemeral bandwidth on core paths includes
the purchased steady up-path bandwidth, as explained in the following.

\emph{(2) Ephemeral bandwidth on core links.} Let $sBW_{S}$ be the
total amount of steady bandwidth sold by a core $AS_S$ for \emph{all}
steady paths in $AS_S$'s ISD. Let $sBW_u$ be the reserved bandwidth
sold for a \emph{particular} steady up-path $S_u$ in this ISD. Let
further $sBW_C$ be the control traffic bandwidth of a core path $C$
between the core ASes of the steady paths for $S$ and $D$. Then,
ephemeral reservations on $C$ launched via $S_u$ can be up to
\begin{equation}
  \label{eq:eph:core}
  eBW_{uC} = \frac{sBW_u}{sBW_S} \cdot \bwrat \cdot sBW_C
\end{equation}
In other words, the ephemeral bandwidth reservable on $C$ launched via steady
path $S_u$ depends not only on the amount of total ephemeral
bandwidth on $C$, but also on $S_u$'s steady up-path bandwidth in
relation to the total amount of steady up-path bandwidth purchased in
$S_u$'s ISD.

\emph{(3) Ephemeral bandwidth in the destination ISD.} In the
destination ISD, the weighted fair sharing is slightly more complex,
but follows the ideas of the previous cases: the weighting includes
the steady bandwidth of all steady up-paths and all steady
down-paths, as well as the ratios of the bandwidth of the core
contracts. Before explaining the details, we note that the reason for
including also the steady down-paths is to give the destination \AD
control over the minimum amount of traffic it receives along
ephemeral paths.

More precisely, an ephemeral path launched over steady up-path $S_u$
and steady down-path $S_d$ with core path $C$ in between obtains
ephemeral bandwidth
\begin{equation}
  \label{eq:eph:destination}
  eBW_{ud} = \frac{C_{S\to D}}{C_{*\to D}} \cdot \frac{sBW_u}{sBW_S} \cdot \bwrat \cdot sBW_{d}
\end{equation}
where $C_{S\to D}$ is the bandwidth negotiated in the core contract
for $C$ between the core ASes of $S$ and $D$, and $C_{*\to D}$ is the total
amount of bandwidth negotiated in all core contracts between any core
AS and $D$'s core AS.

\autoref{eq:eph:destination} looks similar to
\autoref{eq:eph:core}, with an additional factor in the weighting
that reflects the ratio of incoming traffic from other core ASes.
Intuitively, this factor assures that traffic from every other core
AS obtains its fair share based on the bandwidth negotiated in the
individual bilateral contracts.

Finally, the overall bandwidth for an ephemeral path
between end hosts $S$ and $D$ launched over
steady up-path $S_u$ reads
\begin{equation}
  eBW_{uCd} = min(eBW_u, eBW_{uC}, eBW_{ud})
\end{equation}

These equations compute the guaranteed bandwidth using the envisioned long-term
ratio of 5\% steady traffic, 80\% ephemeral traffic, and 15\% best-effort
traffic. Ideally, the ratio should be adjustable by each AS, initially with an
imbalance in favor of best-effort during incremental deployment of \name, until the
number of \name subscribers increases. The overall bandwidth $eBW_{uCd}$ that
can be obtained during early deployment is the minimum of the individual ratios
for each AS and their link bandwidth. We discuss the choice of the ratio in
Section~\ref{sec:parameters} and its adaption in terms of an incremental
deployment strategy in Section~\ref{sec:deployment}.

\paragraph{Fair sharing of steady paths} A challenging question is whether a
fair sharing mechanism is necessary for steady bandwidth. A steady up-path is
used solely by the \AD that requested it, and its use is monitored by the \AD,
which splits the steady up-path bandwidth between its end hosts. In contrast,
steady down-paths need to be revealed to several potential source \ADs, either
as \emph{private} steady down-paths (e.g., for a company's internal services),
or as \emph{public} steady down-paths (e.g., for public services). To
prevent a botnet residing in malicious source \ADs from flooding steady
down-paths, \name uses a weighted fair sharing scheme similar to ephemeral
paths: each \AD using a steady down-path obtains a fair share proportional to
its steady up-path, and its ISD's core path. We give the details of the scheme
in Appendix~\ref{appendix:sharing-down-paths}.

\paragraph{Efficient bandwidth usage via statistical multiplexing}
Internet traffic often exhibits a cyclical pattern, with alternating levels of
utilized bandwidth.
In situations of low utilization, fixed
allocations of bandwidth for steady and ephemeral paths that are
unused would result in a waste of bandwidth. \name reduces such
bandwidth waste through statistical multiplexing,
i.e., unused steady and ephemeral bandwidth is temporarily given to
best-effort flows. A small slack of unallocated steady and ephemeral
bandwidth still remains to accommodate new steady and ephemeral
bandwidth requests. As more and more entities demand steady paths and
their fair share of ephemeral paths, \name gradually squeezes
best-effort flows and releases the borrowed steady and ephemeral
bandwidth up to the default allocations.

\paragraph{Renewal} End hosts can launch ephemeral path renewals to increase
the reserved bandwidth and extend the expiration time of the ephemeral path.
Since ephemeral reservations have a short lifetime, they are frequently
renewed. Renewals are launched using the old reservation which contains the
bandwidth class of the reservation; therefore routers can rapidly decide on the
fastpath how much bandwidth they should allocate for the renewal, for instance
if the bandwidth increased, decreased, or remained the same. Reservations are
given a $\mathit{reservation}\ \mathit{index}$, incremented for each renewal of a specific
ephemeral path. Reservations can be renewed anytime before they expire, and the
end host is allowed to switch to the newer reservation at any time. However,
the end host is not allowed to use both the old and the renewed reservation at
the same time; \autoref{sec:mon_renew} shows a mechanism to detect such
misbehavior.

\subsection{Priority traffic monitoring and policing}
\label{sec:mon}

\noindent Flows that violate their reservations may undermine the
guarantees of other legitimate flows. An ideal monitoring algorithm
should immediately catch \emph{every} such malicious flow. This,
however, would be too expensive for line-rate traffic in the Internet
core.

Instead, as the first line of defense, \name relies on edge \ADs to perform fine-grained traffic
monitoring.  Edge \ADs rely on flow IDs to check each flow's bandwidth usage
and compare it against the reserved bandwidth for that flow ID, which is stored
by each \AD locally during the reservation request.  Previous research has
shown that per-flow slowpath operations are feasible at the edge of the
network~\cite{Stoica2003}.

Monitoring on transit \ADs, however, needs to be processed on the fastpath.  To
detect misbehaving \ADs that purposely fail to regulate their own traffic,
\name deploys a lightweight monitoring mechanism in transit \ADs. First, each
\AD monitors the bandwidth usage of incoming traffic from each neighbor \AD and
compares it against the \emph{total} bandwidth reserved for that neighbor. Such
coarse-grained monitoring timely detects a misbehaving neighbor that
failed to correctly police its traffic.

\begin{figure}[t]
  \begin{center}
    \includegraphics[width=0.9\linewidth]{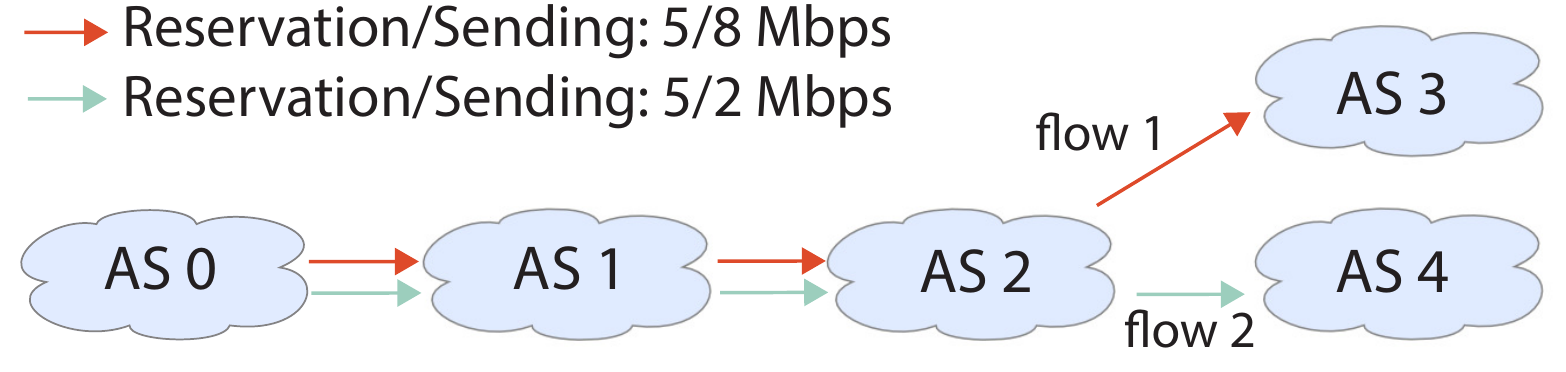}
  \end{center}
  \vspace{-3mm}
  \caption{Per-neighbor monitoring may label benign $AS_1$ malicious.}
  \vspace{-7mm}
  \label{fig:monitoring}
\end{figure}

\paragraph{Why per-neighbor monitoring is insufficient} There are cases,
though, when per-neighbor monitoring in transit \ADs is insufficient.
\autoref{fig:monitoring} depicts two flows originating in $AS_0$, each having
reserved $5$ Mbps. Flow 1 is malicious and sends traffic with $8$ Mbps, while
flow 2 underuses its reservation. $AS_0$ hence does not properly monitor its
flows.  When $AS_1$ performs per-neighbor monitoring, it can only notice that,
in the aggregate, it receives $10$ Mbps from $AS_0$ and sends $10$ Mbps to $AS_2$.
However, when the two flows diverge, $AS_2$ detects flow 1 as malicious and
holds $AS_1$ responsible, although $AS_1$ properly performed per-neighbor
monitoring. 

For this reason, \name additionally utilizes \textit{fine-grained} probabilistic
monitoring of individual flows at the transit \ADs, using a a recently proposed
technique~\cite{hao}. Each transit \AD monitors, per given time interval, all
the flows in a number of randomly chosen bandwidth classes.  Recall that the
bandwidth class of a flow is authenticated by the RTs in the packet header.  In
case the average bandwidth utilization of a flow during that time interval
exceeds the flow's bandwidth class, the flow is classified as malicious and
added to a blacklist, preventing its renewal.

We emphasize that transit \ADs perform monitoring on the fastpath.  Only in
case of suspicion of misbehavior, these \ADs perform out-of-band slowpath
monitoring to police misbehaving neighbors and flows.

To localize the origin \AD of the malicious flow, an \AD informs the previous
\AD of the misbehaving flow. In response, the previous \AD can simply monitor
that specific flow explicitly.  If the violation persists, the suspicious
previous-hop neighbor is likely to be malicious. Then, the \AD can punish it,
for instance, by terminating their contract.

\subsection{Flow renewal monitoring and policing}\label{sec:monitoring-policing}
\label{sec:mon_renew}

\noindent
A successful ephemeral path renewal replaces the old
reservation, therefore the renewal receives the same flow ID as the old
reservation.  However, \name paths allow for RTs with
overlapping validity periods. Therefore, if multiple renewals occur before the
ephemeral path expires, the source would be in possession of multiple sets of
valid RTs: Some corresponding to the ephemeral path with the previous bandwidth class
and old expiration time, and the others corresponding to the new values for
bandwidth class and expiration time, along the same path.  Since all sets of
RTs are associated with the same flow ID, routers would overwrite their
per-flow entries with the new bandwidth class.

A malicious end host could thus exploit renewals by using both sets of RTs, old
and new, during the overlap time of the RTs, thus using more bandwidth than the
amount reserved. To prevent such misuse, end hosts are not allowed to use old
RTs after having used the renewed RTs. When renewals use the same bandwidth
class as the old reservation, simultaneous use of old and new RTs is detected
by the per-class monitoring mechanism (as described above) since the usage is
jointly accounted under the same flow ID.

We now consider the case when the renewed bandwidth class is different from the
old one. The edge AS performs per-flow stateful inspection and is supposed to
filter out traffic that violates the sending rule. Therefore, the edge \AD can
be held accountable by other \ADs for improperly filtering traffic. In transit
\ADs, however, we propose a probabilistic approach for catching this type of
misbehavior. \ADs maintain one Bloom filter~\cite{bloom} per currently active
expiration time and per bandwidth class. Since an RT is maximally valid for 16
seconds and the time granularity is 4 seconds, 4 Bloom filters are needed per
bandwidth class, to record flow IDs that use the bandwidth class within that
time period. We will discuss the details about time discretization in \name in
\autoref{subsec:request}. For an incoming packet with a reservation in a
monitored class $C$, \ADs simply store the tuple $\langle \mathit{flow\ ID},
\mathit{reservation\ index}\rangle$ in the Bloom filter of class $C$. By
probabilistically inspecting some of these Bloom filters, each \AD notices
whether a flow ID uses two different bandwidth classes during a time period.

We further optimize the monitoring algorithm as follows. \name selects a small
number of classes to monitor at a given moment in time, therefore \ADs store
Bloom filters only for the few monitored traffic classes. In addition, \name
does not investigate all Bloom filters: We observe that, when the renewed
bandwidth is much higher or much lower than the previous bandwidth, using both
the old and new reservations would incur an insignificant bandwidth overuse.
Therefore, if a certain reservation index is used in class $C$, \name
investigates only the Bloom filters of the classes whose bandwidth values are
comparable to $C$'s bandwidth (the comparability of classes is discussed in
\autoref{subsec:request}).  \name investigates whether in these Bloom filters
an index $\mathit{reservation\ index} + i$ is present, where $i \in \{0, 1, \ldots,
15\}$ chosen randomly ($i=0$ detects whether the end host maliciously reuses
the same reservation index).  If found, \ADs increment a violation counter for
the source of that flow ID.  The violation counter allows for Bloom filter
false positives. When the violation counter exceeds a threshold, an alarm is
raised for that sender.  Therefore, the more packets an attacker sends, the
higher the probability of detection.  The policing push back technique can then
localize the source \AD of the misbehaving flow. 

\subsection{Dealing with failures}

\noindent
While bandwidth guarantees along fixed network paths allow for a
scalable design, link failures can still disrupt these paths and thus render
the reservations futile. In fact, leaf \ADs and end hosts are rather
interested in obtaining a bandwidth guarantee than
obtaining a specific network path for their traffic.

\name deals with link failures using two mechanisms: (1)
a failure \emph{detection} technique to remove reservations along
faulty paths, and (2) a failure \emph{tolerance} technique to provide
guarantees in the presence of failures. For (1), \name uses short
expiration times for reservations and keep-alive mechanisms. Steady
paths expire within 3 minutes of creation, but leaf \ADs can extend
the steady paths' lifetime using keep-alive messages. Ephemeral paths
have a default lifetime of 16 seconds, which can be extended by
source end hosts through renewals. Unless keep-alive messages or
renewals are used, reservations are removed from the system within
their default expiration time. By construction, a new reservation
cannot be created on top of faulty paths. For (2), \name allows
leaf \ADs to register multiple disjoint steady paths. We also
envision source end hosts being able to choose a bandwidth
reservation service with high reliability, which would use a
small number of disjoint ephemeral paths to the same destination.

\subsection{Dynamic Interdomain Leased Lines}

\noindent Businesses use \emph{leased lines} to achieve highly reliable
communication links. ISPs implement leased lines virtually through reserved
resources on existing networks, or physically through dedicated network links.
Leased lines are very costly, can require weeks to be set up, and are
challenging to establish across several ISPs.

A natural desire is to achieve properties similar to traditional leased lines,
but more efficiently. GEANT offers a service called ``Bandwidth-On-Demand''
(BoD), which is implemented through the InterDomain Controller
Protocol~\cite{idcp} to perform resource allocations across the participating
providers~\cite{geant:bod}.  Although BoD is a promising step, the allocations
are still heavy-weight and require per-flow state.

With \name's properties, ISPs can offer light\-weight Dynamic Interdomain
Leased Lines (DILLs). A DILL can be composed by two longer-lived steady paths,
connected through a core path, or dynamically set up with an ephemeral path
that is constantly renewed. Thanks to the light\-weight operation of \name,
DILLs can be set up with an RTT setup message and are immediately usable. Our
discussions with operators of availability-critical services have shown that
the DILL model has sparked high interest among operators.

To enable long-term DILLs, valid on the order of weeks, the concept of
ephemeral paths in \name could be reframed: long-term DILLs could use the same
techniques for monitoring and policing as ephemeral paths, however, they would
also introduce new challenges.  To enable long-term DILLs, ISPs need to ensure
bandwidth availability even when DILLs are not actively used, as opposed to
ephemeral bandwidth, which can be temporarily used by best-effort flows.  For
this purpose, ISPs could allocate a percentage of their link bandwidth for
DILLs, besides steady, ephemeral, and best-effort paths. Additionally, for
availability in the face of link failures, ISPs would need to consider active
failover mechanisms. For instance, in architectures that provide path choice,
ISPs could leverage disjoint multipath reservations concentrated in a highly
available DILL.  A detailed design though is out of scope for this paper.

\section{Implementation}
\label{sec:imple}\label{sec:implementation}
\noindent
We present the implementation of senders and routers to launch a
reservation request and to use a reservation. We rely on efficient
data structures and algorithms that enable fastpath processing in the
common case and explain the infrequent operations when \name needs
slowpath processing.

\subsection{Bandwidth reservation setup}\label{subsec:request}

\paragraph{Sender implementation}
A reservation request initiator specifies the following configuration
parameters: a flow ID ($128$ bits), a reservation expiration time ($16$ bits),
bandwidth classes for forward and/or reverse directions ($5$ bits each), a path
direction type ($2$ bits), and a reservation index ($4$ bits). \name considers
time at a granularity of $4$ seconds (which we call \textit{\name seconds}). By
default, steady paths thus have an initial lifetime of $45$ \name seconds, and
ephemeral paths of $4$ \name seconds; nevertheless, these paths can be renewed
at any time. All reservations start at the request time.

We chose \name's bandwidth classes to cover a meaningful range for steady and
ephemeral traffic: there are $12$ steady bandwidth classes according to the
formula $16\cdot{\sqrt{2^i}}$~kbps, where $i \in \{0,1,\ldots,11\}$, ranging
from 16~kbps to $\sim$724~kbps; and $20$ ephemeral bandwidth classes according
to the formula $256\cdot \sqrt{2^i}$~kbps, where $i \in \{0,1,\ldots,19\}$,
ranging from 256~kbps to $\sim$185~Mbps. The exponential growth allows for a
fine-grained allocation of smaller bandwidth values, but more coarse-grained
allocation of larger bandwidth values. Additionally, it enables efficient
monitoring of flow renewals, with a small number of classes having comparable
bandwidth.

The path direction type is a flag that indicates, for a $\langle
\mathit{requester},\mathit{destination} \rangle$ pair, either a uni-directional
reservation, for traffic either sent or received by the requester; or
bi-directional, for traffic sent and received by the requester. The reservation
index is a number specific to a flow, incremented every time the reservation
corresponding to the flow is renewed.

\paragraph{Bandwidth reservation and accounting} To efficiently
manage and account for bandwidth reservations, \name routers maintain
the following data structures: (1) a \emph{bandwidth table}, i.e., an
array of size $k$ storing the currently reserved bandwidth for each
of the router's $k$ neighbors; (2) an \emph{accounting table}, i.e.,
a table with tuples containing the flow ID of a reservation, the
expiration time, the bandwidth class, and the neighbor to/from whom the
reservation is specified; (3) a \emph{pending table}, i.e., a table
(of similar structure as the accounting table) that stores pending
reservations. A reservation is said to be \emph{pending} if it has
been requested, but not used for data transmission. A reservation
with flow ID $i$ is said to be \emph{active} when data has been
transmitted using $i$, i.e., the router has seen $i$ in a data
packet. A reservation for $i$ is said to be \emph{expired} if the
router has not seen packets containing $i$ within a time frame of
$\ell$ \name seconds (details below).

To decide whether a requested amount $\mathit{bw}_r$ can be reserved,
routers perform admission control by comparing $\mathit{bw}_r$ with
the entry in the bandwidth table for the specified
neighbor.\footnote{The reason for considering only the current amount
of available bandwidth when making the admission decision is
justified by the monotonicity of reservations: reservations can never
be set up to start in the future, hence, in the next \name second,
there cannot be less bandwidth available than in the current \name
second (unless new reservations are requested).} In case sufficient
bandwidth is available, the request's flow ID, the expiration time, the
request's bandwidth class, and the neighbor are added to the pending
table. The requested amount $\mathit{bw}_r$ is also added to the
respective entry in the bandwidth table. Yet, at this point, the
router does not add information about the request to the accounting
table. The reason is that the request may fail at a later point, in
which case the accounting table update would have to be reverted. In
a periodic background process, the router checks whether there are
entries in the pending table older than $300$ milliseconds
(sufficient to allow for an Internet round trip
time\footnote{http://www.caida.org/research/performance/rtt/walrus0202
 }). Such entries are considered failed reservations, and thus they
are deleted from the pending table and the corresponding reserved
bandwidth is freed and updated in the bandwidth table.

If the router sees a data packet with flow ID $i$ for the first time,
it implies that the reservation for flow ID $i$ was accepted by all
routers on the path. The reservation becomes active and the entry
with flow ID $i$ is then removed from the pending table and added to
the accounting table.

To periodically reclaim unused ephemeral bandwidth of expired
reservations, a router periodically removes the amount of expired
bandwidth from the bandwidth table. The expiration parameter $\ell$
(e.g., $1\le\ell\le 5$) specifies the lifetime (in \name seconds) of
pending reservations. In order to keep reservations active (even if
no data is transmitted), a source simply sends a keep-alive message
within $\ell$ \name seconds. In a periodic background process, the
router then iterates over the accounting table's entries that
correspond to the last $\ell$ \name seconds. More specifically, the
router checks whether the listed flow IDs occur in a Bloom filter
that is filled while forwarding data packets: to enable fastpath
operation, the flow ID of each incoming data packet is stored in a
Bloom filter, not in the accounting table. Bandwidth reclaim is then
processed in the slowpath.

\paragraph{Intermediate \AD implementation}
The MAC operation of RTs are implemented using CBC-MAC based
on AES. Our AES implementation uses AESni~\cite{aesni}, a fast instruction set available on
Intel and AMD CPUs, which requires only 4.15 cycles per byte to encrypt a 1 kB
buffer in CBC mode.  The key necessary for the MAC operation is
expanded once at the \AD and then used for all RTs generated by that \AD.  
\name uses 32 bits for MACs, which constitutes an optimization,
yet provides sufficient security: a forgery will be detected
with probability $1-2^{-32}$.

During a reservation request, the header for the positive admission of a flow
contains the request configuration values set by the sender and the list of RTs
generated so far.  A field \texttt{Hops} is used to locate the correct offset
for a newly generated RT. In addition, a field \texttt{Extension Flag}
indicates the request path type (bi-/uni-directional), the request status
(successful or failed), and whether the packet carries a reservation request or
a reservation confirmation.

When a request does not pass the admission control, then the corresponding
router sets the extension flag to failed, marks its own \AD in the
\texttt{Decline AS*} field, and resets \texttt{Hops} to zero. Starting with
this \AD, every subsequent \AD on the path towards \rep adds a
\texttt{Bandwidth Offer} field with the offered bandwidth.

We implemented \name on top of a SCION-enabled network, which provides path
control. Our \name implementation provides end-host support through a
\name-enabled gateway, which contains modules for reservation requests and
their confirmation, SCION encapsulation, decapsulation, and a \textit{traffic
hijacking} module. The last element is implemented via NetFilter
Queue~\cite{nfqueue}, and it allows to tunnel legacy IP traffic to a remote
host through the \name-enabled SCION network. Such a design provides \name's
benefits to legacy software, as well as facilitates \name's deployment.

The \name packet header contains SCION-relevant information, such as src/dst
addresses, forwarding path as opaque fields (OFs), the current OF/\marker
indicator, and an optional extension field in which \name's reservation request
messages are encoded. We implemented \name in SCION using \textit{extension
headers}.

\section{Evaluation}
\label{sec:eval}\label{sec:evaluation}
\subsection{Processing on router}

\noindent
We first evaluated \name with respect to the \textit{processing overhead on
routers}.  For our evaluation, we used a traffic generator that initiated
bandwidth reservation requests, and sent
traffic within existing reservations.  The traffic generator was connected to a
software router that performed admission control of the request packets, RT
verification, monitoring for the existent reservations, and then forwarded the
packets. 
Every experiment was conducted 1\,000 times.  
We considered routers placed in both edge and core \ADs, however 
processing time only differed for monitoring operations.
All the tests were conducted on a PC with an Intel Intel Xeon E5-2680 2.7 GHz and
16 GB of RAM, running Linux (64-bit).

First, we investigated the time required by a router to process the \name reservation
request. The average time to process a reservation request was
9.1 $\mu$s, resulting in about 109\,890 that can be processed per second.

Then, we tested the speed of the data packet processing. To this end, we used our
high-performance implementation that deploys Intel's DPDK
framework\footnote{\url{http://dpdk.org/}} for networking operations,
and the AESni extension for cryptographic operations. We set the packet length to 1\,500
bytes.  We measured the time of \name processing (i.e., packet parsing and RT
verification).  It took 0.040 $\mu$s on average to process a single packet, thus
a router is capable
to process about 25 million data packets per second. 
(Note that these times do not include interactions with the NIC).

Next, we investigated the performance of monitoring in the core for two
scenarios: 1 and 100 attackers.  The average processing time was 11.24
$\mu$s for a single attacker, and 9.91 $\mu$s for 100 attackers.
As the results show, the average processing time decreases with an increasing number
of attackers, as blacklisted flows are processed faster.

\begin{figure*}[tb]
   \centering     
   \subfigure[]{\label{fig:intradoc}\includegraphics[width=.65\columnwidth]{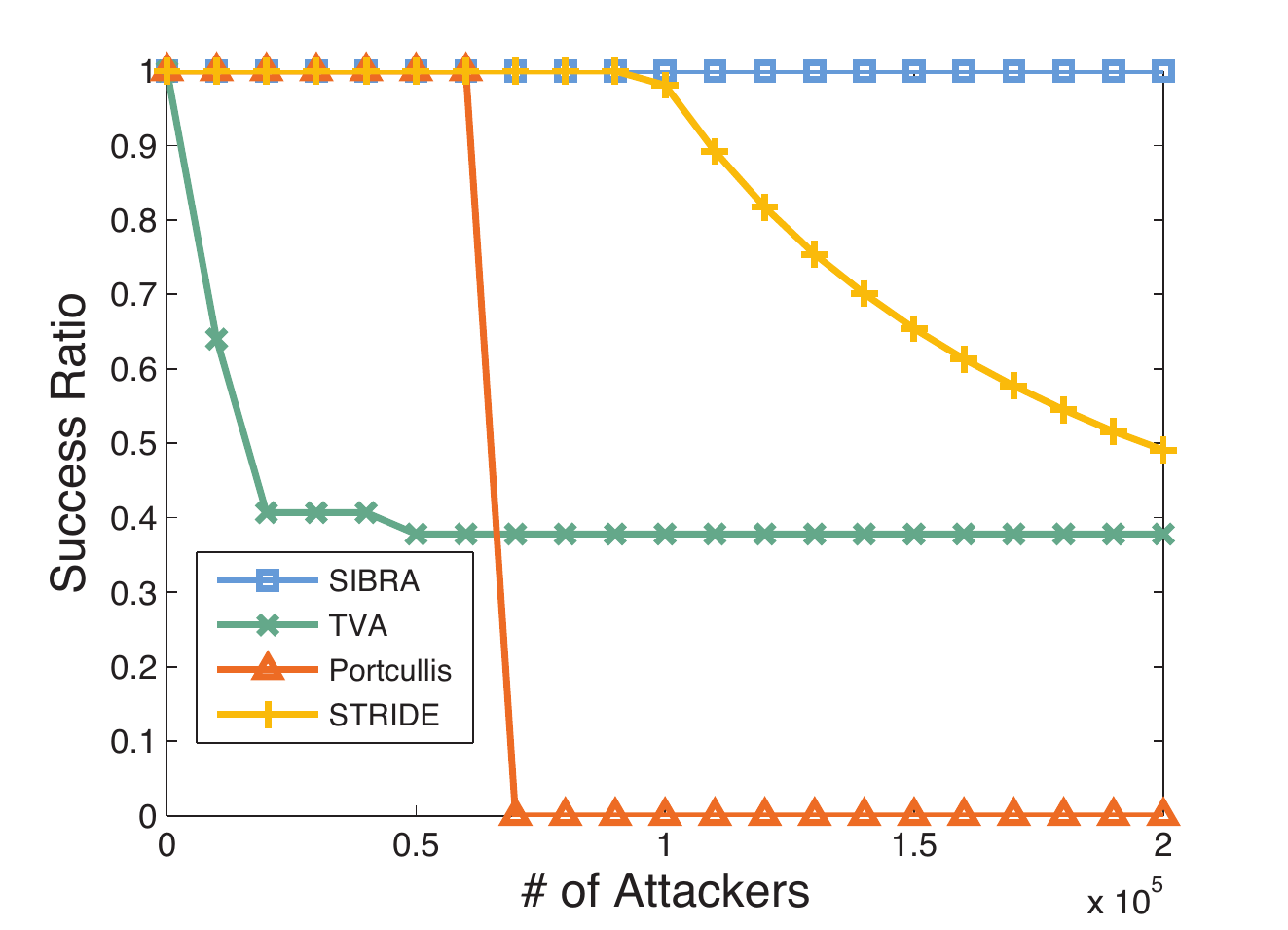}}
   \subfigure[]{\label{fig:interdoc}\includegraphics[width=.65\columnwidth]{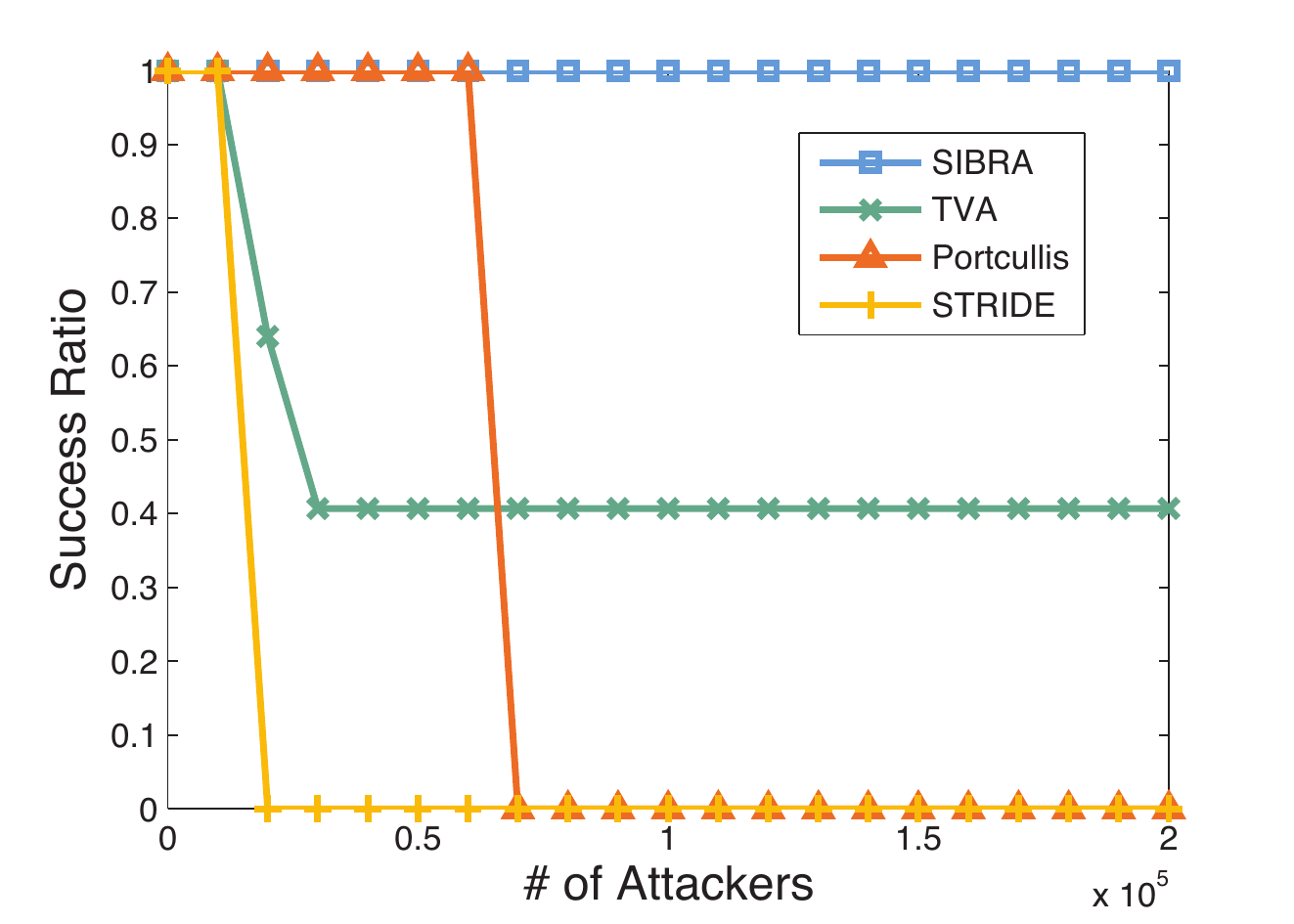}}
   \subfigure[]{\label{fig:coremelt}\includegraphics[width=.65\columnwidth]{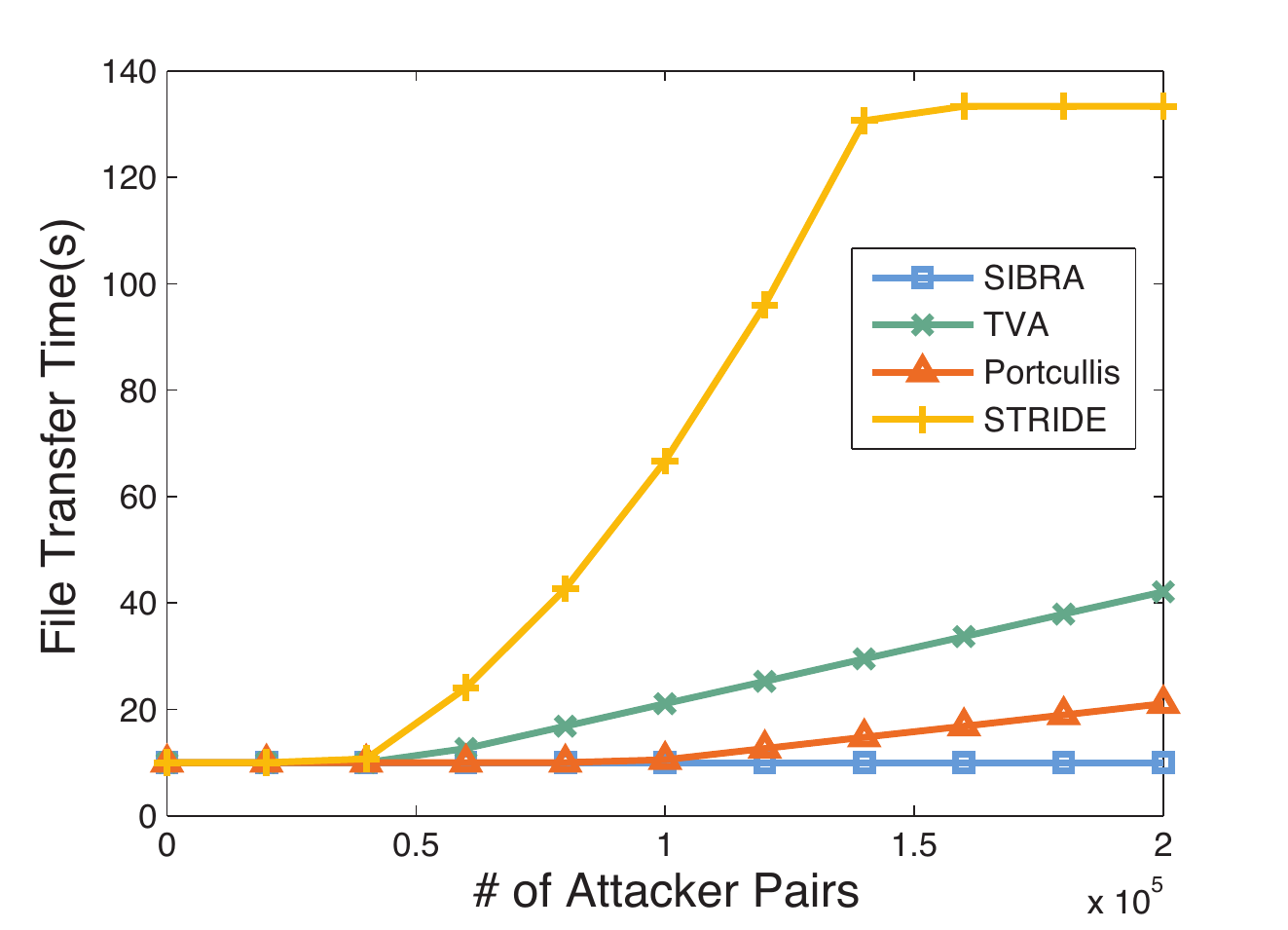}}
   \vspace{-3mm}
   \caption{Comparative simulation results for TVA, Portcullis, STRIDE, and
   \name against Intra-\ISD DoC attack~\ref{fig:intradoc}, Inter-\ISD DoC
attack~\ref{fig:interdoc} and Coremelt attack~\ref{fig:coremelt}.}
\vspace{-6mm}
\end{figure*}

\subsection{Bandwidth guarantees under botnet attacks}\label{subsec:com}

\noindent
To show \name's resilience to Denial of Capability (DoC) and Coremelt attacks,
we run a simulation on an Internet-scale topology. In our simulation, the
attackers attempt to exhaust the bandwidth of the links common with legitimate
flows.  We compare our results with TVA~\cite{yang2005tva},
Portcullis~\cite{Parno2007}, and STRIDE~\cite{hsiao2013stride}, obtained using
the same configuration.

\paragraph{Method} Our Internet-scale topology is based on a CAIDA
dataset \cite{caida} that contains 49\,752 \ADs and the links among them as
observed from
today's Internet. Based on these connections, we grouped the \ADs into five \ISDs,
representing five continent-based regions.  For our simulation we chose the two
biggest \ISDs: $ISD_1$ containing 21\,619, and $ISD_2$ containing 6\,039
\ADs. The core of each \ISD is formed by Tier-1 ISPs.
We set the capacity of the core link between $ISD_1$ and $ISD_2$ to 40 Gbps.
Inside each \ISD, we set the capacity of core links to 10 Gbps, the capacity of
links between a core \AD and a Tier-2 \AD to 2.4 Gbps, and all other links to
640 Mbps.  Steady paths and core paths were established before the experiment. 

In both attack scenarios, the attackers (compromised hosts) are distributed
uniformly at random in different \ADs.  
Legitimate sources reside in two \ADs (i.e., each \AD contains 100 legitimate
sources). 
We further use the same parameters as the related work: a
5\% rate limit for reservation requests, and request packets of 125 bytes.
All the sources (including attackers) send 10 requests per second.  According
to Mirkovic et al.~\cite{mirkovic2009test}, we set 4 seconds as the request
timeout. 

\paragraph{DoC Attack} We simulate both intra-\ISD and inter-\ISD DoC attacks.
For the intra-\ISD case, source and destination \ADs are within $ISD_2$, and $ISD_2$
contains 1\,000 contaminated \ADs. All the requests, from benign and
malicious \ADs, traverse the same link in the core. 
In the inter-\ISD scenario, the source resides in $ISD_1$ and the destination resides in
$ISD_2$, there are 500 contaminated \ADs in each \ISD, and all
the requests traverse the same links in the core.

Figures~\ref{fig:intradoc} and~\ref{fig:interdoc} show the fraction of
successfully delivered capability requests (\textit{success ratio}) correlated
to the number of active attackers. For both cases (intra- and inter-\ISD DoC
attacks), TVA and Portcullis perform similarly: on core links, legitimate
requests mingle with malicious ones. Afterwards, since the link bandwidth
decreases after traversing the core, there is a rapid increase in the request
packets' queueing time.  Consequently, the success ratio decreases. TVA's
success ratio stabilizes around 40\%.  Portcullis uses computational puzzles,
and the request packets with a higher computational level are forwarded first.
Hence, when more attackers with optimal strategy~\cite{Parno2007} appear, the
time to compute a puzzle increases accordingly, leading to a decrease of the
success ratio to 0 when the computation time exceeds 4 seconds. In STRIDE, the
\ISD core has no protection, but traffic inside $ISD_2$ has a higher priority
than traffic coming from $ISD_1$. Thus, during the intra-\ISD attack, STRIDE's
success ratio stays 100\% until the core becomes congested.  However, in the
inter-\ISD case, STRIDE's performance declines dramatically, since a majority
of requests from $ISD_1$ are dropped if any core link in $ISD_2$ is congested.
\name successfully delivers all the legitimate requests, in both attack
scenarios, because \name requests are launched using steady paths, and steady
paths guarantee a fair share of control traffic along core paths.

\paragraph{Coremelt Attack} We simulate a Coremelt attack with the following
settings: $ISD_2$ contains 500 pairs of contaminated \ADs (selected uniformly at
random),
which communicate using ephemeral paths, each with a throughput
of 8~kbps of their 256~kbps reservations. 
The source and the destination also communicate
using an ephemeral path, of 800~kbps. All the ephemeral paths in the experiment
traverse the same core link.  We measure the bandwidth obtained when the source
sends to the destination a 1 MB file.

Figure~\ref{fig:coremelt} shows that the congestion on the core link degrades
the file transfer time in STRIDE to over 100 seconds.  TVA, which uses
per-destination queues to forward authorized traffic, performs slightly worse
than Portcullis, simulated using per-source weighted fair sharing based on the
computational level. \name outperforms the other schemes, because it gives a
lower bound on the bandwidth obtained for the file transfer, due to its
weighted fair sharing based on the steady paths.

\begin{figure}[t]
   \centering
   \includegraphics[width=\columnwidth]{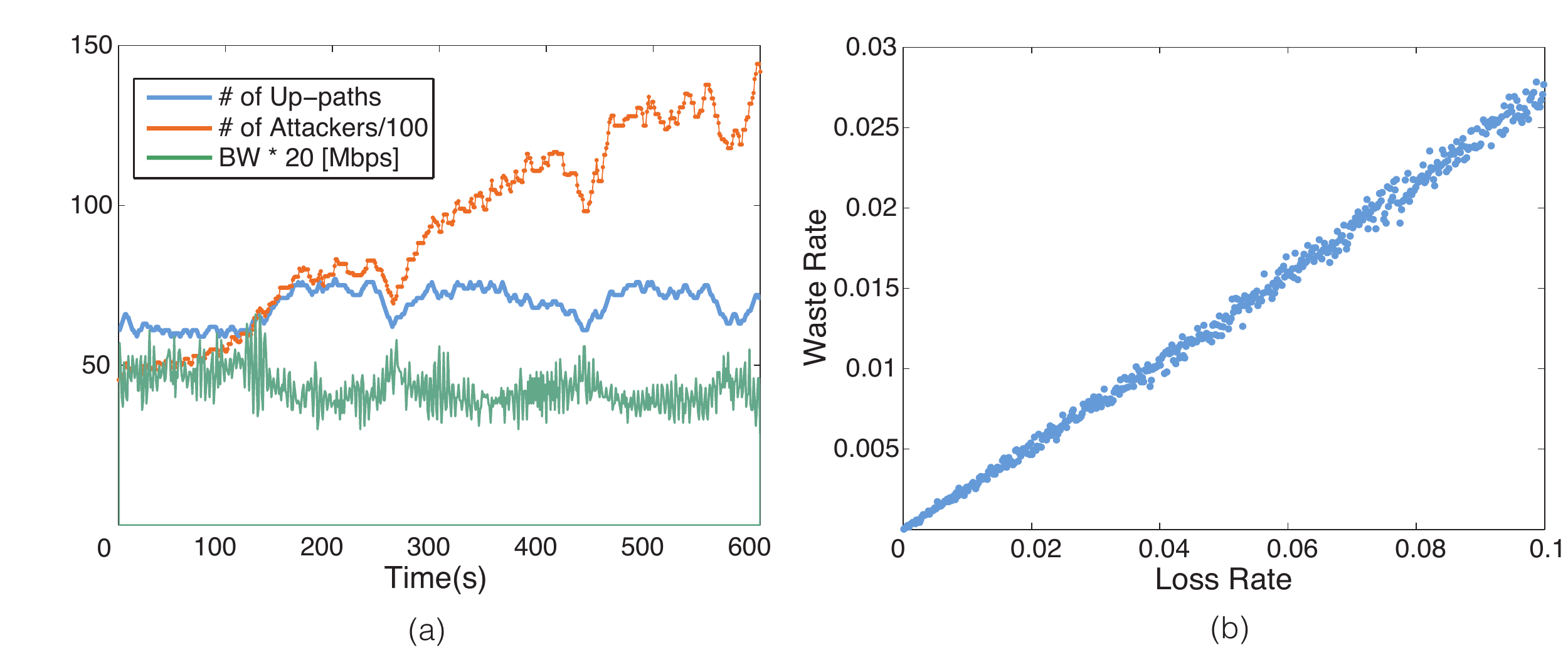}
   \vspace{-5mm}
   \caption{Simulation results on \name's availability. (a) shows
   the existence of the reservable bound for bandwidth requests. Note that the
   bandwidth (green line) in the figure is multiplied by 20 for improved readability.
   (b) presents the resilience of bandwidth reservation against
   packet loss.}
   \vspace{-7mm}
\label{fig:mix}
\end{figure}

\subsection{Lower bound on bandwidth fair share}\label{subsec:bound}

\noindent
We simulate the bandwidth obtained by new ephemeral paths when requests for
ephemeral paths arrive from both benign and malicious sources.  We considered a
scenario where all the requests are forwarded using the same steady down-path
(\name's worst case for weighted fair sharing).

The legitimate steady up-path from the source \AD carried 5 requests per
second, and has a bandwidth of 362~kbps.  There were approximately 50 attackers
on every malicious up-path, and each attacker sent one request per second. The
attackers' steady up-path bandwidth was randomly selected from our steady
bandwidth classes (16~kbps to 724~kbps).  The bandwidth requested for
ephemeral paths ranged from 256~kbps up to 11.6~Mbps.

The result for this setting is presented in Figure~\ref{fig:mix}(a). The green
line shows the real-time reservable bandwidth, that changes dynamically but
finally stabilizes around 2.5 Mbps. At time interval 100, the number of
attackers and steady up-paths used for requesting ephemeral paths increases.
However, \name guarantees that reservable bandwidth remains stable despite the
increasing numbers of attackers.  This is due to the fair share, which is not
affected by the number of attackers with steady paths.

\subsection{Reservation request loss tolerance}

\noindent
Next, we simulate the influence of packet loss on ephemeral bandwidth reservation. We
assume that at every second there are 1\,000 reservation requests sent, with the
following parameters: variable path length (5--10), random bandwidth
(50~kbps -- 6.4~Mbps), variable
packet loss rate (0--10\%), and RTT set to 1 second. Similar to
Portcullis~\cite{Parno2007} and TVA~\cite{yang2005tva}, we assume that request
packets are limited to 5\% of the entire link capacity.

In our simulation, we consider packet loss for both reservation request and
reply packets.  This setting introduced unused bandwidth reservation on the
routers that had already processed the packet, until bandwidth reclaim occurs.
We express the bandwidth waste rate $r_{\mathit{waste}}$ as unused reserved
bandwidth divided by the sum of reserved bandwidth.

As shown in Figure~\ref{fig:mix}(b), even at a loss rate of 5\%, the
corresponding $r_{\mathit{waste}}$ is no more than 1.4\%.  Moreover, the diagram
indicates that $r_{\mathit{waste}}$ increases linearly when the loss rate rises,
which shows that \name tolerates packet loss well, thus providing robust
bandwidth reservation.

\section{Incremental Deployment}
\noindent
Within a single ISP network, deployment of \name does not require major changes
in the underlying infrastructure since the ISP can utilize its existing core
network with protocol-independent transport like MPLS. The ISP can thus build a
``\name-ready'' network by adding new customer/provider edge routers and
setting up MPLS tunnels with reserved bandwidth among them to traverse the
traditional network fabric. A global-scale inter-ISP deployment is
more challenging, because a simple overlay approach with IP-tunneling would not
provide the contiguous bandwidth reservation required for \name. To take full
advantage of \name, ISPs need direct links to interconnect their \name routers.
Therefore, in its initial deployment phase, we envision a \name network
operated by a small group of ISPs with mutual connectivity.

An essential question is whether such a partially-deployed new network
infrastructure provides immediate financial benefits for early adopters, and
subsequently attracts new ISPs. The business example of the startup company
Aryaka is similar to \name regarding the deployment purposes. Aryaka has
successfully established a private core network infrastructure, dedicated to
optimize WAN traffic between Aryaka's Points of Presence (POPs) across the
world. These POPs deploy Aryaka's proprietary WAN optimization protocols, and
enterprise customers' distributed business sites located near POPs benefit from
application acceleration. By offering a global network solution, Aryaka gained
the interest of regional ISPs that want to provide WAN optimization beyond
their own regions. Aryaka is continuously expanding its edge infrastructure
through Tier-3 and Tier-4 ISPs. Yet, as opposed to \name, by using a private
core network, Aryaka's solution comes at a high cost, and may be even more
costly to scale to all ASes in the Internet.

Similar to the case of Aryaka, we expect \name's deployment to begin at the
core, between a few Tier-1 ISPs that seek to provide DILLs spanning their joint
regions. These early adopters may quickly monetize the \name bandwidth
reservation service by selling DILLs to their direct customers. Gradually, the
\name network would expand through new ISP collaborators interested in
providing bandwidth reservation beyond their own regions. ISPs have the
incentive to support \name, as they can draw traffic towards them, and also
appeal to both existing and new clients who desire effective DDoS protection,
thus increasing the ISPs' revenues.

During the expansion of \name, ISPs are likely to start \name deployment with
lower ratios for steady and ephemeral bandwidth, suitable for the needs of a
small number of initial \name customers.  Meanwhile, best-effort customers
still enjoy a throughput similar to that before \name deployment. As the number
of \name subscribers increases, ISPs could locally adjust the ratios towards
an increased steady and ephemeral proportion, and
persuade their providers to follow, as well as adjust their core contracts
accordingly. As more and more customers shift from best-effort to
\name, best-effort traffic obtains a smaller ratio.
Depending on their customer segmentation, ISPs could either adjust best-effort
subscriptions to the new network traffic, or increase their link capacity.

\begin{figure}[t]
   \centering     
   \includegraphics[width=.45\columnwidth]{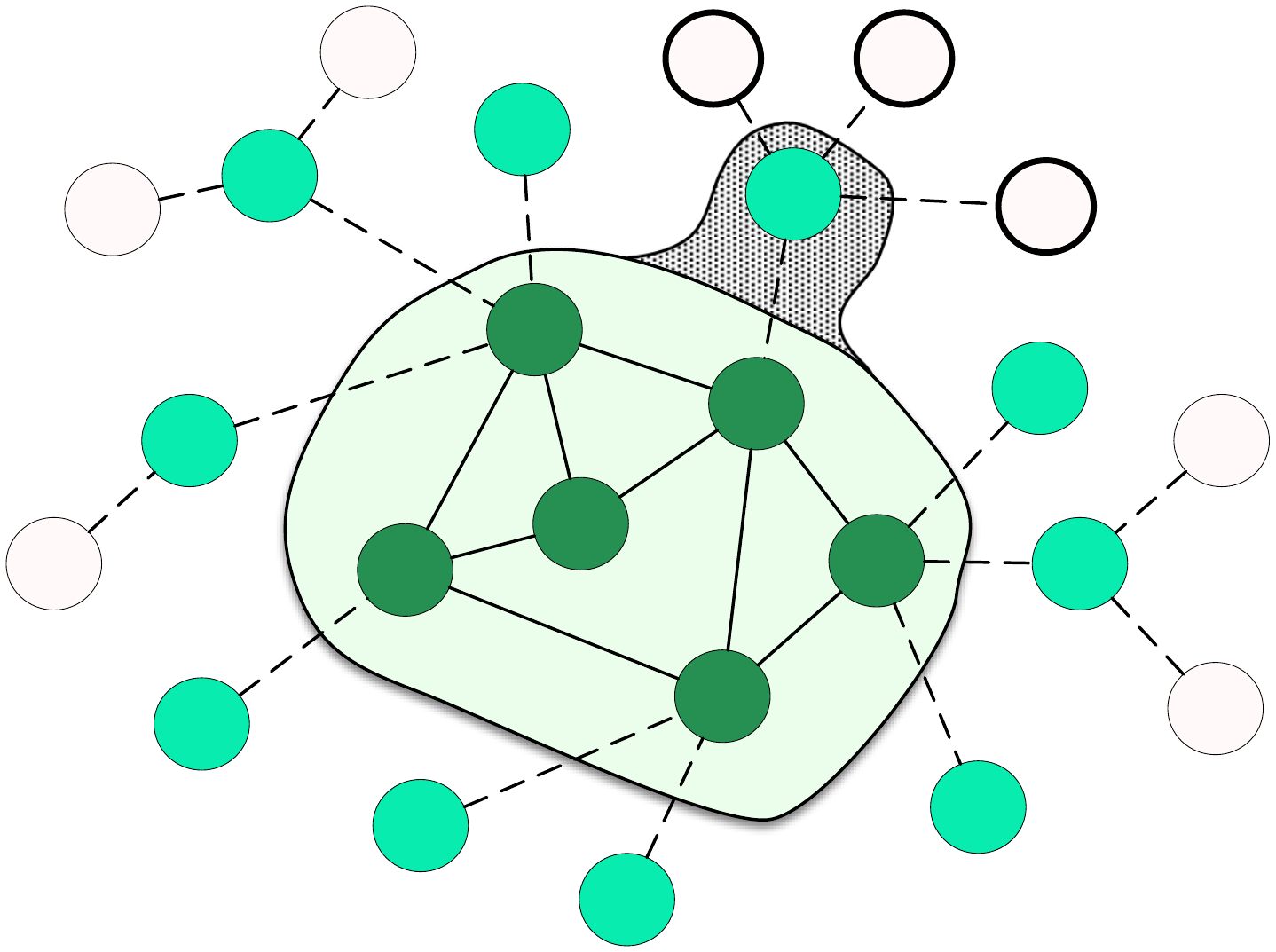}
   \caption{Deploying ISPs (dark colors) gain revenue from all
their neighbors (medium colors) potentially buying guaranteed bandwidth.
The deploying region extends through neighbors (patterned area), with their
direct neighbors as potential buyers (bold outline).}
\vspace{-5mm}
\label{fig:deployment}
\end{figure}

We evaluated a potential deployment plan for \name using the AS topology from
CAIDA\footnote{\label{CAIDA}\url{http://www.caida.org/data/as-relationships/}}
in the following setting.  We considered a set of initial adopters, tier-1 ISPs
selected uniformly at random. Potential adopters in the next deployment round
are the neighbors of the deploying nodes, as depicted in
~\autoref{fig:deployment}, such that there is always a contiguous region of
deploying ASes. We consider rational potential adopters, which deploy \name
only if they can monetize the guaranteed-bandwidth service by selling it to
their neighbors. Such neighbors would buy the service if the traffic they
originate can use DILLs up to their destinations. Thus, we compare the traffic
originating at a buyer neighbor AS that can use DILLs, compared to the total
amount of traffic originating at the same neighbor AS. Since traffic information between
ASes is usually confidential, we approximate the traffic using a model
introduced by Chan et al.~\cite{chan2006adoptability}: the traffic between a
source and a destination AS is represented by the product of the ASes' IP
spaces. We obtained the data on the AS-IP-space mapping from
CAIDA\footnote{\url{http://data.caida.org/datasets/routing/routeviews-prefix2as/}}.

When the set of initial deployers consists of three ASes, next round adopters
could monetize \name on a percentage of traffic between 40\% -- 48\%.
Four initial adopters lead to potential \name traffic of 47\% -- 49\%,
and five initial adopters to 50\% -- 52\%. We conclude that deployment
starting at the Internet core greatly leverages the incremental deployment of
\name.

\label{sec:deployment}

\section{Use Cases}
\noindent
With the flexible lifetime of DILLs, ranging from tens of seconds to weeks
on-demand, \name brings immediate benefits to applications where
guaranteed availability matters. These applications comprise
critical infrastructures, such as financial services and smart electric grids,
as well as business applications, such as videoconferencing and reliable data
sharing in health care.
As discussed above, setting up leased lines in these cases may take several
weeks and may become prohibitively expensive:
it is costly to install leased lines between each pair of domains, and also to
connect each domain through a leased line to a central location in order to
build up a star topology.

\paragraph{Critical infrastructures} Financial services, for instance
\textit{transaction processing from payment terminals}, would become more
reliable when using \name DILLs: since DILLs guarantee availability even
in the presence of adversarial traffic, payment requests and their confirmations
would \emph{always} obtain a guaranteed minimum bandwidth.
DILLs could also be used for \textit{remote monitoring of
power grids}: a minimum guaranteed bandwidth would be suitable to deliver the
monitored parameters, independent of malicious hosts exchanging traffic.
\textit{Telemedicine} is another use case of practical
relevance: the technology uses telecommunication to provide remote health care
--- often in critical cases or emergency situations where interruptions
could have fatal consequences.

\paragraph{Business-critical applications} \textit{Videoconferencing} between
the remote sites of a company receives increasing importance as a convenient
way to foster collaborations while reducing travel costs. Short-lived and
easily installable DILLs provide the necessary guaranteed on-demand bandwidth
for reliably exchanging video traffic. Another application is \textit{reliable
on-demand sharing of biomedical data} for big-data processing, complementing
the efforts of improving health care quality and cost in initiatives such as
\emph{Big Data to Knowledge} launched by the US National Institutes of Health
(NIH)~\cite{health}.

\section{Discussion}
\label{sec:discussion}\label{sec:discuss}

\subsection{On the choice of bandwidth proportions for \name links}
\label{sec:parameters}

\noindent
Recall that in \autoref{sec:overview}, we assigned 80\%, 15\%, and 5\% of a
link's bandwidth to ephemeral, best-effort, and steady paths, respectively.
This parameter choice is justified through an analysis of today's actual
Internet traffic.

\begin{compactitem}

  \item First to notice is that the majority of traffic constitutes persistent
  high-bandwidth connections: for example in Australia, we see that Netflix's
  video connections contribute to more than 50\% of the entire Internet traffic
  \cite{netflix-au}. Given an additional amount of traffic from other large
  video providers such as Youtube and Facebook, we estimate ephemeral paths to
  require roughly 70--90\% of a link's bandwidth.

  \item Best-effort is still important for some types of low-bandwidth
  connections:
  email, news, and SSH traffic could continue as best-effort traffic, totaling
  3.69\% of the Internet traffic~\cite{labovitz2010};
  similar the case for DNS traffic totaling 0.17\% of the Internet
  traffic~\cite{labovitz2010}.
  In addition, very short-lived flows (that is flows with a lifetime less than
  256~ms) with very few packets (the median flow contains 37
  packets~\cite{trammell2012}) are unlikely to establish \name reservations,
  simply to avoid the round-trip time of the reservation setup. Such flows sum
  up to 5.6\% of the Internet traffic~\cite{trammell2012} and can thus also be
  categorized under best-effort.

  \item Finally, regarding the amount of bandwidth for steady paths and
  connection-establishment traffic, we conducted an experiment using the
  inter-\AD traffic summary by a DDoS detection system at one of the largest
  \mbox{tier-1} ISPs. With a 10-day recording of this data, we found that only
  $0.5\%$ of the $1.724\times 10^{13}$ packets were connection establishment
  packets. To enable communication guarantees for low-bandwidth traffic,
  including bandwidth reservation request packets, we designed \name to
  allocate ten-fold of the amount measured.

\end{compactitem}

Since it is hard to specify the actual bandwidth proportions precisely, we use
80\%, 15\%, and 5\% as initial values and note that these values can be
re-adjusted at any point in the future.

We recall from \autoref{sec:ephemeral} that, in addition to the parameter
choice, \name's statistical multiplexing between the traffic classes helps to
dynamically balance the traffic. We expect that in particular the long-lived
reservations are not always fully utilized, in which case best-effort traffic
can be transmitted instead.

\subsection{Per-flow stateless operations are necessary}
\label{sec:stateless}

\noindent To understand the amount of per-flow storage state required on the
fastpath, we investigate the number of active flows per second as seen by a
core router in today's Internet. We used anonymized one-hour Internet traces
from CAIDA, collected in July 2014. The traces contain all the packets that
traversed a 10~Gbps Internet core link of a Tier-1 ISP in the United States,
between San Jose and Los Angeles.

\begin{figure}[tb]
   \centering     
   \includegraphics[width=.8\columnwidth]{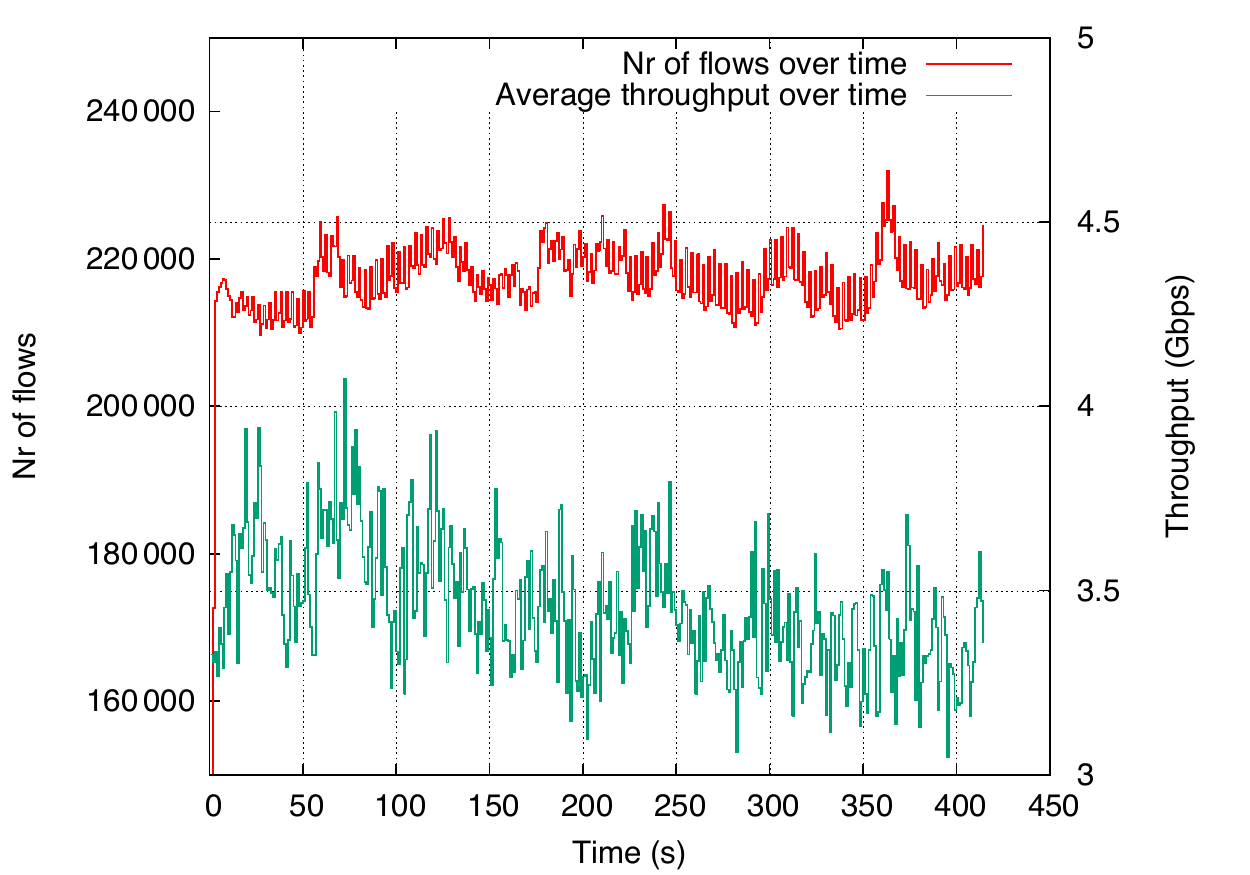}
   \caption{The number of active flows every second and their throughput,
	observed on a 10~Gbps Internet core link.} 
\vspace{-6mm}
\label{fig:1second}
\end{figure}

\autoref{fig:1second} depicts our findings as the number of active flows on the
core link at a granularity of one second, for a total duration of 412 seconds.
We observe that the number of flows varies around 220\,000, with a
boundary effect at the beginning of the data set. These flows sum into a
throughput between 3 and 4~Gbps --- a link load of 30\% to 40\%. A large core
router switching 1~Tbps (with 100 such 10~Gbps links) would thus observe
22$\times10^6$ flows per second \textit{in the normal case}, considering a link
load of only 40\%. \textit{In an attack case}, adversaries could greatly
inflate the number of flows by launching connections between bots, as in
Coremelt~\cite{studer2009coremelt}. Schuchard et al.\ already analyzed attacks
that can exhaust the router memory~\cite{schuchard2010losing}. All these
results suggest storing per-flow state in the fastpath, on the line card, becomes
prohibitively expensive, even more so when the core link load increases.

\subsection{Case study: achievable ephemeral bandwidth on core links}
\label{sec:case}

\noindent A central point of \name is to guarantee a sufficient amount of
bandwidth using today's infrastructure, even for reservations that span
multiple ISDs. A central question is how much bandwidth an end-domain could
minimally obtain if globally all domains attempt to obtain their maximum fair
share. To investigate this point, we considered a scenario with Australia as
destination, and all non-Australian leaf ASes in the world reserving ephemeral
bandwidth to Australia. We picked Australia because with its 24 million
inhabitants, it represents a major economy, and it already experienced
infrastructure congestion in today's Internet~\cite{netflix-au}. While its
geographical location hinders laying new cables, Australia is well-suited for
our study aiming to determine a lower bound on the amount of bandwidth \name
core links can expect. Other countries, especially those situated on larger
continents, typically feature higher-bandwidth connectivity, as laying cables
on land is easier than in the ocean.

\begin{figure}[tb]
   \centering     
   \includegraphics[width=.8\columnwidth]{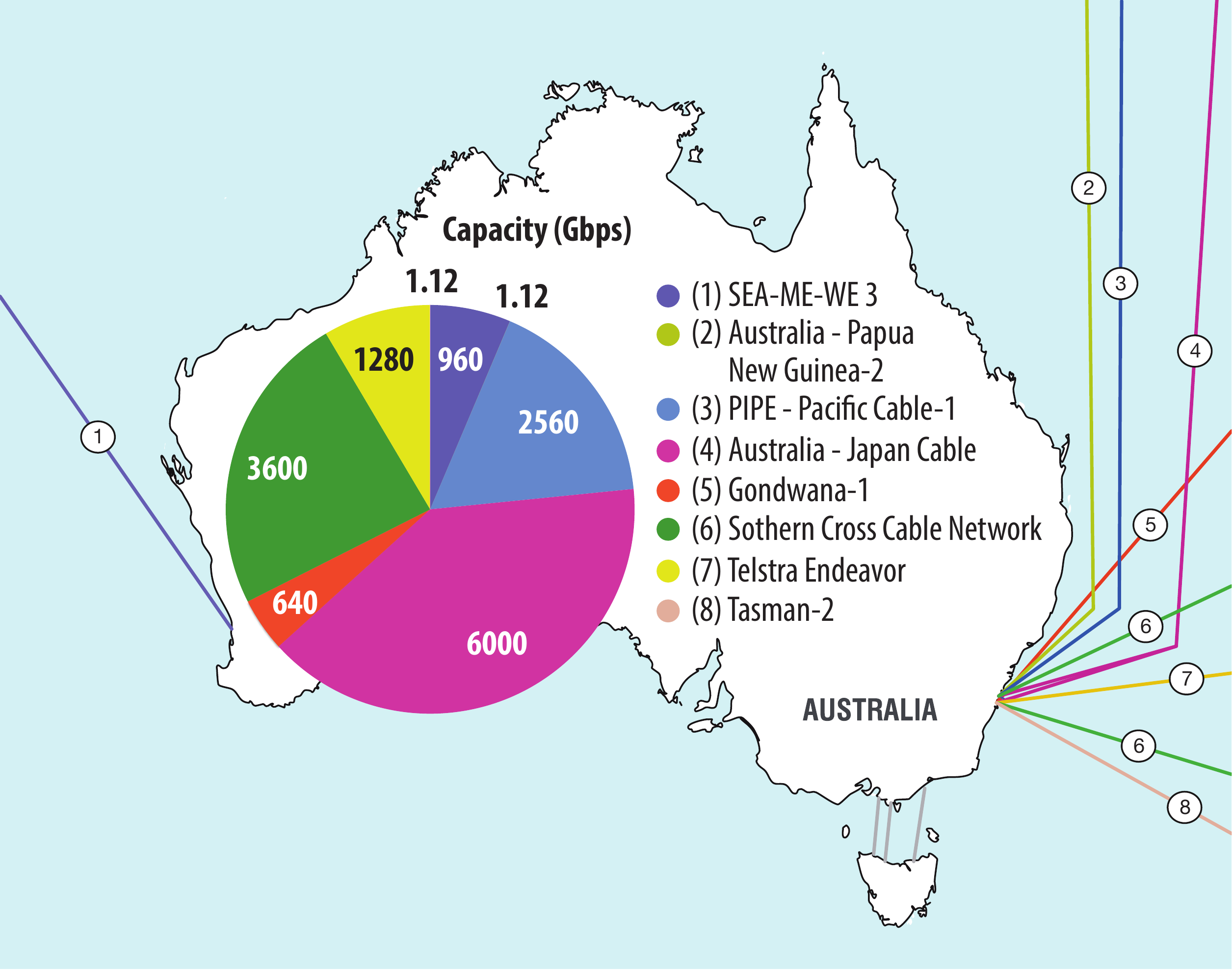}
   \caption{Australia submarine link map, including link capacities.}
\vspace{-6mm}
\label{fig:mapAU}
\end{figure}

~\autoref{fig:mapAU} illustrates the current submarine link map of Australia,
including the name and capacity of the
links.\footnote{\url{http://www.submarinecablemap.com/} illustrates the
submarine link map. The link capacities were obtained from various resources,
e.g., the Australia-Japan Cable capacity from
\url{http://www.ajcable.com/company-history/}.} The entire traffic traverses these
links. 
For simplicity, we assume
guaranteed bandwidth is split equally between leaf ASes.  In practice, however,
the bandwidth is proportional to the size of the steady paths of the leaf ASes
(Section~\ref{sec:path}). We considered two cases: (i) the worst case, i.e.,
when all reservations are squeezed over the same link --- in our case, we chose
the highest-bandwidth cable, namely the Australia-Japan Cable (6~Tbps), and
(ii) the best case, i.e., when the reservations are distributed across all
cables (totaling 15.04~Tbps). In
contrast to other architectures, \name's underlying architecture, SCION, 
enables the use of multi-path
communication for the traffic between a source and a destination, along several
core links.

We have determined the number of leaf ASes in the world, using the AS topology
from CAIDA\footref{CAIDA}, and counted 32\,428 non-Australian leaf ASes using
the AS number and
location\footnote{\url{http://data.caida.org/datasets/as-organizations/}}.
After the analysis, we found that each non-Australian leaf AS obtains a fair
share of (i) 185.02~Mbps (148~Mbps for ephemeral traffic), or (ii) 463.86~Mbps
(371.08~Mbps for ephemeral traffic). We thus conclude that \name's fair sharing
scheme offers a substantial amount of bandwidth through an efficient use of the
current Internet infrastructure. In case this amount is insufficient, an AS
could purchase additional bandwidth for a specific destination from its core
AS.

The prospects are even brighter: considering the planned undersea physical
infrastructure development, the capacity of the cables connecting Australia
with the rest of the world would increase by 168~Tbps by the beginning of 2018.
With such an increase, the fair share on \name's core links becomes 5.64~Gbps
per leaf AS in case (ii).

\section{Related Work}
\label{sec:related}

\noindent
\textbf{Capability-based mechanisms}~\cite{Anderson2004,Yaar2004, yang2005tva,Parno2007,natu2007fine,
lee2010floc,hsiao2013stride} aim at isolating legitimate flows from malicious
DDoS attack traffic.  Network capabilities are access tokens issued by on-path
entities (e.g., routers and \dst) to the \src. Only packets carrying such
network capabilities are allowed to use a privileged channel.  Capability-based
schemes, however, require additional defense mechanisms against Denial of
Capability attacks~\cite{Argyraki2005} and against attacks with colluding hosts
or legitimate-looking bots~\cite{studer2009coremelt,kang2013crossfire}.
To address DoC attacks, \textbf{TVA}~\cite{yang2005tva} tags each packet with a
path identifier which is based on the ingress interface of the traversing ASes.
The path identifier is used to perform fair queueing of the request packets at
the routers. However, sources residing further away from the congested link will
suffer a significant disadvantage.
\textbf{Portcullis} \cite{Parno2007} deploys computational puzzles to provide
per-compu\-tation fair sharing of the request channel. Such proof-of-work
schemes, however, are too expensive to protect every data packet. Moreover,
Portcullis does not provide the property of botnet-size independence.
\textbf{Floc}~\cite{lee2010floc} fair-shares link bandwidth of individual flows
and differentiates between legitimate and attack flows for a given link.
However, such coarse-grained per-AS fair sharing may not always be effective;
in particular, low-rate attack flows can often not be precisely differentiated.
\textbf{CoDef}~\cite{lee2013codef} is a collaborative defense mechanism in
which a congested AS asks the source ASes to limit their bandwidth to a
specific upper bound and to use a specific path. Source ASes that continue
sending flows that exceed their requested quota are classified as malicious.
CoDef does not prevent congestion in the first place, but instead retroactively
handles one congested link at a time. Since congestion can still occur
on links, sources cannot be given a guarantee for reaching a destination.
\textbf{STRIDE}~\cite{hsiao2013stride} is a capability-based DDoS protection
architecture that builds on several concepts from SCION~\cite{zhang2011scion,scion2015}.
Although STRIDE shares similarities with \name (steady paths and ephemeral
paths), STRIDE lacks intra-core and inter-ISD communication guarantees;
STRIDE's intra-domain guarantees are built on the assumption of
congestion-free core networks. Moreover, STRIDE lacks monitoring and policing
mechanisms, as well as an implementation.

\paragraph{Resource allocation}
Several queuing protocols~\cite{Shreedhar1996, Pan2000, Stoica2003}
have been proposed to approximate fair bandwidth allocation at routers.
Their correctness, however, relies on the trustworthiness of the routers and
flow identifiers.
The Path Computation Element (\textbf{PCE})
architecture~\cite{ash2006path,vasseur2009path} computes inter-\AD
routes and enables resource allocation across \AD boundaries in
Generalized Multi-Protocol Label Switching (GMPLS) Traffic Engineered
networks. However, the discovery of inter-\AD PCE path fragments
discloses information about other cooperating \AD, such as the internal
topology. Some \ADs will be reluctant to share this information due
to confidentiality reasons.

\paragraph{Resource reservation}
\textbf{RSVP}~\cite{Zhang1993} is a signaling protocol for bandwidth
reservation. Because RSVP is not designed with security in mind, the
reservation may fail due to DDoS attacks. RSVP requires the sender
(e.g., a host or an \AD when RSVP aggregation is used as specified in
RFC 3175) to make an end-to-end reservation to the receiver(s),
causing a quadratic number of control messages (in the number of
entities) in the network and quadratic state on the intermediate
routers.

\section{Conclusions}
\label{sec:conclusion}

\noindent Through hierarchical decomposition of resource reservations,
\name is the first scalable architecture that provides \emph{inter-domain} bandwidth
guarantees --- achieving botnet-size independence and resolving even sophisticated
DDoS attacks such as Core\-melt~\cite{studer2009coremelt} and Cross\-fire~\cite{kang2013crossfire}.
\name ends the arms race between DDoS attackers and defenders, as it
provides guaranteed resource reservations regardless of the
attacker's botnet size.
A salient property of \name is that it can be built without requiring
per-flow state in the fastpath of a router, resulting in a simple
router design and high-speed packet processing.
We anticipate that \name becomes a game changer in the battle
against large-scale DDoS attacks.

\section{Acknowledgments}

\noindent
We would like to thank Virgil Gligor, Chris Pappas, Christian Rossow,
Stephen Shirley, and Laurent Vanbever for insightful discussions and their valuable
comments throughout the evolution of this project. We also thank Xiaoyou
Wang, Dominik Roos, and Takayuki Sasaki for their help with the implementation and evaluation of \name.

The research leading to these results has received funding from the European
Research Council under the European Union's Seventh Framework Programme
(FP7/2007-2013) / ERC grant agreement 617605.  We also gratefully acknowledge
support by ETH Zurich, and NSF under award number CNS-1040801. The research was
also supported by a gift from KDDI.

\bibliographystyle{IEEEtranS}
\bibliography{0-string,bib}

\begin{thebibliography}{10}
\providecommand{\url}[1]{#1}
\csname url@samestyle\endcsname
\providecommand{\newblock}{\relax}
\providecommand{\bibinfo}[2]{#2}
\providecommand{\BIBentrySTDinterwordspacing}{\spaceskip=0pt\relax}
\providecommand{\BIBentryALTinterwordstretchfactor}{4}
\providecommand{\BIBentryALTinterwordspacing}{\spaceskip=\fontdimen2\font plus
\BIBentryALTinterwordstretchfactor\fontdimen3\font minus
  \fontdimen4\font\relax}
\providecommand{\BIBforeignlanguage}[2]{{%
\expandafter\ifx\csname l@#1\endcsname\relax
\typeout{** WARNING: IEEEtranS.bst: No hyphenation pattern has been}%
\typeout{** loaded for the language `#1'. Using the pattern for}%
\typeout{** the default language instead.}%
\else
\language=\csname l@#1\endcsname
\fi
#2}}
\providecommand{\BIBdecl}{\relax}
\BIBdecl

\bibitem{idcp}
``Inter-domain controller ({IDC}) protocol specification,''
  \url{http://www.controlplane.net/idcp-v1.1-ns/idc-protocol-specification-v1.1.pdf},
  2010.

\bibitem{caida}
``{Center for Applied Internet Data Analysis (CAIDA)},''
  \url{http://www.caida.org/home/}, 2014.

\bibitem{netflix-au}
``{Netflix congesting the Australian Internet},''
  \url{http://www.smh.com.au/digital-life/digital-life-news/these-graphs-show-the-impact-netflix-is-having-on-the-australian-internet-20150402-1mdc1i.html},
  2015.

\bibitem{website:nanog}
``{North American Network Operators' Group},''
  \url{https://www.nanog.org/list}, 2015.

\bibitem{website:400}
``{Technical Details Behind a 400Gbps NTP Amplification DDoS Attack},''
  \url{https://blog.cloudflare.com/technical-details-behind-a-400gbps-ntp-amplification-ddos-attack},
  2015.

\bibitem{AIP}
D.~G. Andersen, H.~Balakrishnan, N.~Feamster, T.~Koponen, D.~Moon, and
  S.~Shenker, ``{Accountable Internet Protocol (AIP)},'' in \emph{ACM SIGCOMM},
  2008.

\bibitem{Anderson2004}
T.~Anderson, T.~Roscoe, and D.~Wetherall, ``{Preventing Internet
  Denial-of-Service with Capabilities},'' \emph{ACM SIGCOMM Computer
  Communication Review}, 2004.

\bibitem{Argyraki2005}
K.~Argyraki and D.~R. Cheriton, ``{Network Capabilities: The Good, the Bad and
  the Ugly},'' in \emph{ACM HotNets}, 2005.

\bibitem{scion2015}
D.~Barrera, R.~M. Reischuk, P.~Szalachowski, and A.~Perrig, ``{SCION} five
  years later: Revisiting scalability, control, and isolation on
  next-generation networks,'' \emph{arXiv e-prints}, 2015.

\bibitem{bloom}
B.~H. Bloom, ``Space/time trade-offs in hash coding with allowable errors,''
  \emph{Communications of the ACM}, 1970.

\bibitem{chan2006adoptability}
H.~Chan, D.~Dash, A.~Perrig, and H.~Zhang, ``{Modeling adoptability of secure
  BGP protocol},'' in \emph{ACM SIGCOMM}, 2006.

\bibitem{demers1989fair}
A.~Demers, S.~Keshav, and S.~Shenker, ``Analysis and simulation of a fair
  queueing algorithm,'' \emph{ACM SIGCOMM Comp. Comm. Rev.}, 1989.

\bibitem{ash2006path}
A.~Farrel, J.-P. Vasseur, and J.~Ash, ``A path computation element
  ({PCE})-based architecture,'' Tech. Rep., 2006.

\bibitem{geant:bod}
GEANT, ``Bandwidth on demand,'' \url{http://geant3.archive.geant.net/
  service/BoD/pages/home.aspx}, 2015.

\bibitem{godfrey2009pathlet}
P.~Godfrey, I.~Ganichev, S.~Shenker, and I.~Stoica, ``Pathlet routing,'' in
  \emph{ACM SIGCOMM Comp. Comm. Rev.}, 2009.

\bibitem{aesni}
S.~Gueron, ``{Intel Advanced Encryption Standard (AES) New Instructions Set},''
  Intel, 2010, white paper 323641-001, Revision 3.

\bibitem{hardin1968tragedy}
G.~Hardin, ``The tragedy of the commons,'' \emph{Science}, 1968.

\bibitem{rfc6253}
T.~Heer and S.~Varjonen, ``Host identity protocol certificates,'' 2011.

\bibitem{hsiao2013stride}
H.-C. Hsiao, T.~H.-J. Kim, S.~B. Lee, X.~Zhang, S.~Yoo, V.~Gligor, and
  A.~Perrig, ``{STRIDE}: Sanctuary trail -- refuge from internet {DDoS}
  entrapment,'' in \emph{AsiaCCS}, 2013.

\bibitem{diffserv}
F.~B. J.~Babiarz, K.~Chan, ``{Configuration Guidelines for DiffServ Service
  Classes}.''

\bibitem{kang2013crossfire}
M.~S. Kang, S.~B. Lee, and V.~D. Gligor, ``{The Crossfire Attack},'' in
  \emph{IEEE S\&P}, 2013.

\bibitem{labovitz2010}
C.~Labovitz, S.~Iekel-Johnson, D.~McPherson, J.~Oberheide, and F.~Jahanian,
  ``{Internet Inter-Domain Traffic},'' \emph{ACM SIGCOMM}, 2010.

\bibitem{LasJohChu08:userDirectedRouting}
P.~Laskowski, B.~Johnson, and J.~Chuang, ``User-directed routing: From theory,
  towards practice,'' in \emph{ACM NetEcon}, 2008.

\bibitem{lee2010floc}
S.~B. Lee and V.~D. Gligor, ``{FLoc}: Dependable link access for legitimate
  traffic in flooding attacks,'' in \emph{IEEE ICDCS}, 2010.

\bibitem{lee2013codef}
S.~B. Lee, M.~S. Kang, and V.~D. Gligor, ``{CoDef}: Collaborative defense
  against large-scale link-flooding attacks,'' in \emph{ACM CoNEXT}, 2013.

\bibitem{health}
R.~Margolis, L.~Derr, M.~Dunn, M.~Huerta, J.~Larkin, J.~Sheehan, M.~Guyer, and
  E.~D. Green, ``{The National Institutes of Health's Big Data to Knowledge
  (BD2K) initiative: capitalizing on biomedical big data},'' \emph{Journal of
  the American Medical Informatics Association}, 2014.

\bibitem{mirkovic2009test}
J.~Mirkovic, S.~Fahmy, P.~Reiher, and R.~K. Thomas, ``How to test {DoS}
  defenses,'' in \emph{CATCH}, 2009.

\bibitem{rfc5201}
R.~Moskowitz, P.~Jokela, T.~R.~Henderson, and T.~Heer, ``Host identity protocol
  version 2.''

\bibitem{rfc970}
J.~Nagle, ``{On Packet Switches with Infinite Storage},'' 1985.

\bibitem{natu2007fine}
M.~Natu and J.~Mirkovic, ``Fine-grained capabilities for flooding {DDoS}
  defense using client reputations,'' in \emph{ACM LSAD}, 2007.

\bibitem{Pan2000}
R.~Pan, B.~Prabhakar, and K.~Psounis, ``{CHOKe-a stateless active queue
  management scheme for approximating fair bandwidth allocation},'' in
  \emph{IEEE INFOCOM}, 2000.

\bibitem{Parno2007}
B.~Parno, D.~Wendlandt, E.~Shi, A.~Perrig, B.~Maggs, and Y.-C. Hu,
  ``{Portcullis: Protecting Connection Setup from Denial-of-Capability
  Attacks},'' in \emph{ACM SIGCOMM}, 2007.

\bibitem{raghavan2009secure}
B.~Raghavan, P.~Verkaik, and A.~C. Snoeren, ``Secure and policy-compliant
  source routing,'' \emph{IEEE/ACM Transactions on Networking}, vol.~17, no.~3,
  2009.

\bibitem{schuchard2010losing}
M.~Schuchard, A.~Mohaisen, D.~Foo~Kune, N.~Hopper, Y.~Kim, and E.~Y. Vasserman,
  ``{Losing control of the internet: using the data plane to attack the control
  plane},'' in \emph{ACM CCS}, 2010.

\bibitem{Shreedhar1996}
M.~Shreedhar and G.~Varghese, ``{Efficient fair queuing using deficit
  round-robin},'' \emph{IEEE/ACM Transactions on Networking}, 1996.

\bibitem{stoica2002internet}
I.~Stoica, D.~Adkins, S.~Zhuang, S.~Shenker, and S.~Surana, ``{Internet
  Indirection Infrastructure},'' \emph{ACM SIGCOMM Comp. Comm. Rev.}, 2002.

\bibitem{Stoica2003}
I.~Stoica, S.~Shenker, and H.~Zhang, ``{Core-Stateless Fair Queueing: A
  Scalable Architecture to Approximate Fair Bandwidth Allocations in High-Speed
  Networks},'' \emph{IEEE/ACM Transactions on Networking}, 2003.

\bibitem{studer2009coremelt}
A.~Studer and A.~Perrig, ``The {Coremelt} attack,'' in \emph{ESORICS}, 2009.

\bibitem{trammell2012}
B.~Trammell and D.~Schatzmann, ``{On Flow Concurrency in the Internet and its
  Implications for Capacity Sharing},'' in \emph{ACM CSWS}, 2012.

\bibitem{vasseur2009path}
J.~Vasseur and J.~Le~Roux, ``Path computation element communication protocol,''
  \emph{IETF RFC 5557}, 2009.

\bibitem{nfqueue}
H.~Welte and P.~N. Ayuso, ``{The netfilter.org libnetfilter\_queue project},''
  \url{http://www.netfilter.org/projects/ libnetfilter_queue/}, 2014.

\bibitem{intserv}
J.~Wroclawski, ``{The Use of RSVP with IETF Integrated Services},'' 1997.

\bibitem{hao}
H.~Wu, H.-C. Hsiao, and Y.-C. Hu, ``Efficient large flow detection over
  arbitrary windows: An algorithm exact outside an ambiguity region,'' in
  \emph{ACM IMC}, 2014.

\bibitem{Yaar2004}
A.~Yaar, A.~Perrig, and D.~Song, ``{SIFF: A Stateless Internet Flow Filter to
  Mitigate DDoS Flooding Attacks},'' in \emph{IEEE S\&P}, 2004.

\bibitem{yang2007nira}
X.~Yang, D.~Clark, and A.~W. Berger, ``Nira: A new inter-domain routing
  architecture,'' \emph{IEEE/ACM Transactions on Networking}, 2007.

\bibitem{yang2005tva}
X.~Yang, D.~Wetherall, and T.~Anderson, ``A {DoS}-limiting network
  architecture,'' \emph{ACM SIGCOMM Comp. Comm. Rev.}, 2005.

\bibitem{Zhang1993}
L.~Zhang, S.~Deering, D.~Estrin, S.~Shenker, and D.~Zappala, ``{RSVP: A New
  Resource ReSerVation Protocol},'' \emph{IEEE Network}, 1993.

\bibitem{zhang2011scion}
X.~Zhang, H.-C. Hsiao, G.~Hasker, H.~Chan, A.~Perrig, and D.~G. Andersen,
  ``{SCION: Scalability, Control, and Isolation on Next-generation Networks},''
  in \emph{IEEE S\&P}, 2011.

\end{thebibliography}

\clearpage

\appendix
\renewcommand{\thesubsection}{\Alph{subsection}}
\subsection{Fair sharing of steady down-paths}
\label{appendix:sharing-down-paths}

\noindent Recall from \autoref{sec:corepaths} that core ASes negotiate core
contracts to set up core paths among each other (the double continuous lines in
\autoref{fig:topology}). The reserved bandwidth for those core paths is
negotiated based on aggregated traffic volumes as observed in the past. The
question we consider in the following is how the reserved bandwidth is split
among the customers of the core ASes. More precisely, we describe a sharing
mechanism that assigns each leaf AS $E$ a fair amount of bandwidth for $E$'s
traffic traversing the core paths. Intuitively, \emph{fair} in this context
means proportional to the amount of bandwidth that $E$ has purchased for its
steady up-paths to the core AS.
In contrast to the fair sharing mechanism for ephemeral paths
(Section~\ref{sec:ephemeral}), the equations we introduce here do not require
the additional weighting factor $\bwrat=\frac{\bweph}{\bwstd}$ given by the
ratio of ephemeral and steady bandwidth.

\paragraph{Steady bandwidth on core links}
The steady bandwidth of a core path $\mathbf{C}=\langle AS_{C1}, \ldots,
AS_{Cn}\rangle$ between core $AS_{C1}$ and a destination core $AS_{Cn}$ is
split between all customer ASes of $AS_{C1}$, weighted with the bandwidth of
the steady up-path each customer AS uses.

Let $sBW_{u*}$ be the total amount of steady bandwidth sold by a core $AS_{C1}$
for \emph{all} its steady paths. Let $sBW_u$ be the reserved bandwidth sold by
$AS_{C1}$ for a \emph{particular} steady up-path $S_u$. Let further
$sBW_\mathbf{C}$ be the steady bandwidth of core path $\mathbf{C}$. Then the
steady traffic on $\mathbf{C}$ launched via $S_u$ can be up to:

\begin{equation}
  \label{eq:std:core}
  sBW_{uC} = \frac{sBW_u}{sBW_{u*}} \cdot sBW_\mathbf{C}
\end{equation}

\paragraph{Steady bandwidth in the destination ISD}
In the destination ISD, the weighted fair sharing follows the same intuition,
this time consisting first of a fair sharing mechanism between different ISDs,
then between the ASes in the same ISD. 

More precisely, steady traffic launched over steady up-path $S_u$ and steady
down-path $S_d$, with core path $\mathbf{C}$ in between, obtains a throughput:
\begin{equation}
  \label{eq:std:destination}
  sBW_{uCd} = \frac{C_{C1\to Cn}}{C_{*\to Cn}} \cdot \frac{sBW_u}{sBW_{u*}} \cdot sBW_{d}
\end{equation}
where $C_{C1\to Cn}$ is the bandwidth negotiated in $\mathbf{C}$'s core
contract between core $AS_{C1}$ (source ISD) and core $AS_{Cn}$ (destination
ISD), $C_{*\to Cn}$ is the total amount of bandwidth negotiated in the core
contracts between \emph{any} core \AD and the destination ISD's core $AS_{Cn}$,
and $sBW_d$ is the reserved bandwidth sold by core $AS_{Cn}$ for steady down-path
$d$.

The second case is that of local steady traffic in the destination ISD,
which does not traverse any core path. In fact, this happens only when the
steady up- and down-paths cross at the same core \AD, otherwise traffic would
traverse a core path between the core \ADs of the destination ISD (see
\autoref{fig:topology}).

This case introduces a \emph{preference} $\rho \in(0,1)$ that splits the
bandwidth of the steady down-path between traffic that traverses core paths,
and traffic that does not.

Let $sBW_{u*}$ be the total bandwidth of all steady up-paths to a given core AS
in the destination ISD. Then, a particular steady up-path among them, say with
bandwidth $sBW_{u}$, obtains steady bandwidth on the down-path $d$ of
\begin{equation}
  \label{eq:std:onlyd}
  sBW_{ud}^\rho = \frac{sBW_{u}}{sBW_{u*}} \cdot \rho \cdot sBW_{d}
\end{equation}

Traffic from external ISDs is weighted accordingly, extending
\autoref{eq:std:destination}:
\begin{equation}
  sBW_{uCd}^\rho = (1-\rho) \cdot sBW_{uCd}
\end{equation}

\paragraph{Authentication of bandwidth values} To compute $sBW_{uC}$ and
$sBW_{uCd}$, \ADs use the bandwidth values for steady up-paths and core paths
included in the \name packet headers. In order to prevent a malicious AS from
increasing its fair share by tampering with the bandwidth values, \name
requires the core ASes of each ISD to sign the bandwidth of steady up-paths
inside the ISD, as well as the bandwidth of core paths starting at the ISD core.

Signing can take place when the steady paths are registered at the ISD core.
Each AS first verifies the signatures, and then computes the fair shares. Fast
signing and verification can be implemented using the public-key high-speed
signature scheme \texttt{ed25519}. The private key used for signing by an ISD
core can be shared among the core \ADs in each ISD. For signature verification,
the corresponding public key of an ISD core is distributed to all other ISDs,
which disseminate the key to all \ADs inside the ISD. In practice, on the order
of only 100 public keys need to be distributed. Since ISD cores are typically
stable, key change and re-distribution should be infrequent. A detailed
description is out of scope for this paper.

\paragraph{Dynamic fair sharing}
A pertinent question is whether the computed fair share is large enough to be
useful.  On core links, whose capacity exceeds hundreds of Gbps (recall the
example of Australia in Section~\ref{sec:discussion}), even a ratio of only 1\%
steady bandwidth is on the order of a few Gbps, which would give each AS
outside Australia tens of Mbps of steady traffic.

However, in the destination ISD, given that steady down-paths' capacity is
merely a few Mbps of bandwidth, the fair share obtained for each leaf AS is
less than 1~kbps --- too small to be useful. Of course, a popular destination
\AD could register several steady down-paths to increase the steady fair share,
but the increase factor is still small compared to the number of possible
source \ADs.

The key observation is that \name enforces the fair share on steady traffic
\textit{dynamically}, i.e., only when a steady down-path becomes congested. The
fair share is computed between the \ADs that \textit{actively use} the steady
down-path, as opposed to all possible \ADs in the Internet. The fair share is
reserved for 1 second, which allows the \AD two retry attempts if an ephemeral
reservation fails (considering an RTT of 300 ms -- a conservative value,
according to measurements performed by CAIDA\footnote{\url{https://www.caida.
org/research/performance/rtt/walrus0202/}}).

\end{document}